\documentclass[utf8]{arXivHarvard} 

\usepackage{url,lineno,microtype,subcaption}
\usepackage[onehalfspacing]{setspace}
\usepackage{siunitx}
\pdfoutput=1
\usepackage{caption, subcaption}
\usepackage{dirtytalk}
\usepackage{ulem}

\usepackage{etoolbox}

\usepackage{xcite}

\usepackage{xr}
\makeatletter
\newcommand*{\addFileDependency}[1]{
  \typeout{(#1)}
  \@addtofilelist{#1}
  \IfFileExists{#1}{}{\typeout{No file #1.}}
}
\makeatother

\newcommand*{\myexternaldocument}[1]{%
    \externaldocument{#1}%
    \addFileDependency{#1.tex}%
    \addFileDependency{#1.aux}%
}

\myexternaldocument{SupplementaryMaterial_frontiers}

\def\PC-m{phase-contrast}
\def\FL-m{fluorescent}

\newcommand{\dd}{\textit{D. discoideum}}

\definecolor{ao(english)}{rgb}{0.0, 0.5, 0.0}
\newcommand{\new}[1]{\textcolor{black}{#1}}
\newcommand{\old}[1]{\textcolor{blue}{\sout{#1}}}

\newcommand{\switchOld}[1]{%
  \ifthenelse{\equal{#1}{0}}{\renewcommand{\old}[1]{}}{}}
\switchOld{0}

\newcommand{\switchNew}[1]{%
  \ifthenelse{\equal{#1}{0}}{\renewcommand{\new}[1]{\textcolor{red}{#1}}}{}}
\switchNew{1}

\def\keyFont{\fontsize{8}{11}\helveticabold }
\def\firstAuthorLast{Moldenhawer {et~al.}} 
\def\Authors{Ted Moldenhawer\,$^{1}$, Eduardo Moreno\,$^{2}$, Daniel Schindler\,$^{3}$, Sven Flemming\,$^{1}$, Matthias Holschneider\,$^{3}$, Wilhelm Huisinga\,$^{3}$, Sergio Alonso\,$^{2}$ and Carsten Beta\,$^{1,*}$}
\def\Address{$^{1}$Institute of Physics and Astronomy, University of Potsdam, Potsdam, Germany\\
$^{2}$Department of Physics, Universitat Politècnica de Catalunya, Barcelona, Spain\\
$^3$Institute of Mathematics, University of Potsdam, Potsdam, Germany}

\def\corrAuthor{Carsten Beta}
\def\corrEmail{beta@uni-potsdam.de}

\begin{document}
\onecolumn
\firstpage{1}

\title[Spontaneous transitions between modes of migration]{Spontaneous transitions between amoeboid and keratocyte-like modes of migration} 

\author[\firstAuthorLast ]{\Authors} 
\address{} 
\correspondance{} 

\extraAuth{}

\maketitle

%
%

\begin{abstract}
The motility of adherent eukaryotic cells is driven by \old{coherent patterns of activity in} \new{the dynamics of} the actin cytoskeleton.
Despite the common force-generating actin machinery, different cell types often show diverse modes of locomotion that differ in their shape dynamics, speed, and persistence of motion.
Recently, experiments \old{with the model organism} \new{in} \textit{Dictyostelium discoideum} have revealed that different motility modes can be induced in \old{the same cell type} \new{this model organism}, depending on genetic modifications, developmental conditions, and synthetic changes of intracellular signaling.
Here, we report experimental evidence that \new{in a mutated \dd{} cell line with increased Ras activity,} switches between two distinct migratory modes, the amoeboid and fan-shaped type of locomotion, can \new{even} spontaneously occur within the same cell.
We observed \new{and characterized} repeated and reversible switchings between the two modes of locomotion, suggesting that \new{they} \old{both migratory modes} are \old{stable coexisting} \new{distinct} behavioral \old{states} \new{traits that coexist} within the same cell. \old{with spontaneous noise-induced transitions occurring between them.}
\old{Furthermore,} We adapted an established \new{phenomenological} motility model that combines a reaction-diffusion system for the intracellular dynamics with a dynamic phase field to account for our experimental findings.

\tiny
 \keyFont{ \section{Keywords:} cell migration, amoeboid motility, keratocytle-like motility, modes of migration, \dd{}, actin dynamics} 
\end{abstract}

\section{Introduction}
\label{sec:introduction}
Actin-driven motility is an essential prerequisite for a wide range of biological functions, such as wound healing, immune responses, or embryonic development.
The underlying signaling pathways and cytoskeletal structures may vary widely, depending on the cell type and biological context.
Consequently, the associated cell morphodynamics and speeds of propagation are diverse, resulting in a wide range of different modes of motility.
For example, the persistent locomotion of a fish keratocyte is clearly distinct from the more erratic motion of neutrophils or dendritic cells.
In most cases, different modes of motility are linked to specific cell types and are closely related to their biological functions.
\new{However, there are also cases where migratory plasticity of the same cell type is an essential requirement for cellular function~\citep{yamadaMechanisms3DCell2019,nikolaouStressfulTumourEnvironment2020,friedlTumourcellInvasionMigration2003}.
For example, in order to achieve metastatic dissemination, cancer cells show a wide range of different migration strategies, including single celled amoeboid and mesenchymal motion as well as multicellular streaming and collective cell migration~\citep{friedlCancerInvasionMicroenvironment2011,psailaMetastaticNicheAdapting2009,chambersDisseminationGrowthCancer2002}.
Here, individual cells undergo dramatic transformations to change their adhesion properties, remodel their actin cytoskeleton, and alter the activity of their associated signaling pathways~\citep{yilmazMechanismsMotilityMetastasizing2010a,vignjevicReorganisationDendriticActin2008,bearDirectedMigrationMesenchymal2014a}.}
\old{However, in some cases, also transitions between different migratory modes are observed within the same cell type, for example when cancer metastasis is initiated.}
\new{Despite these insights, many fundamental questions regarding migratory plasticity remain unresolved and call for experiments that rely on simpler model systems.}
\old{Nevertheless, the switching between different styles of migration within the same cell type has received only little attention so far and different migratory modes have mostly been studied in different cell types individually.}
\new{In particular, a} \old{But} better understanding of switches in migratory modes will not only provide functional insights into processes such as cancer metastasis but will also elucidate whether specific dynamical actin structures are characteristics of individual cell types or universally emerge in different cellular contexts.

\begin{figure}[t]
    \centering
    \includegraphics[width=0.7\linewidth]{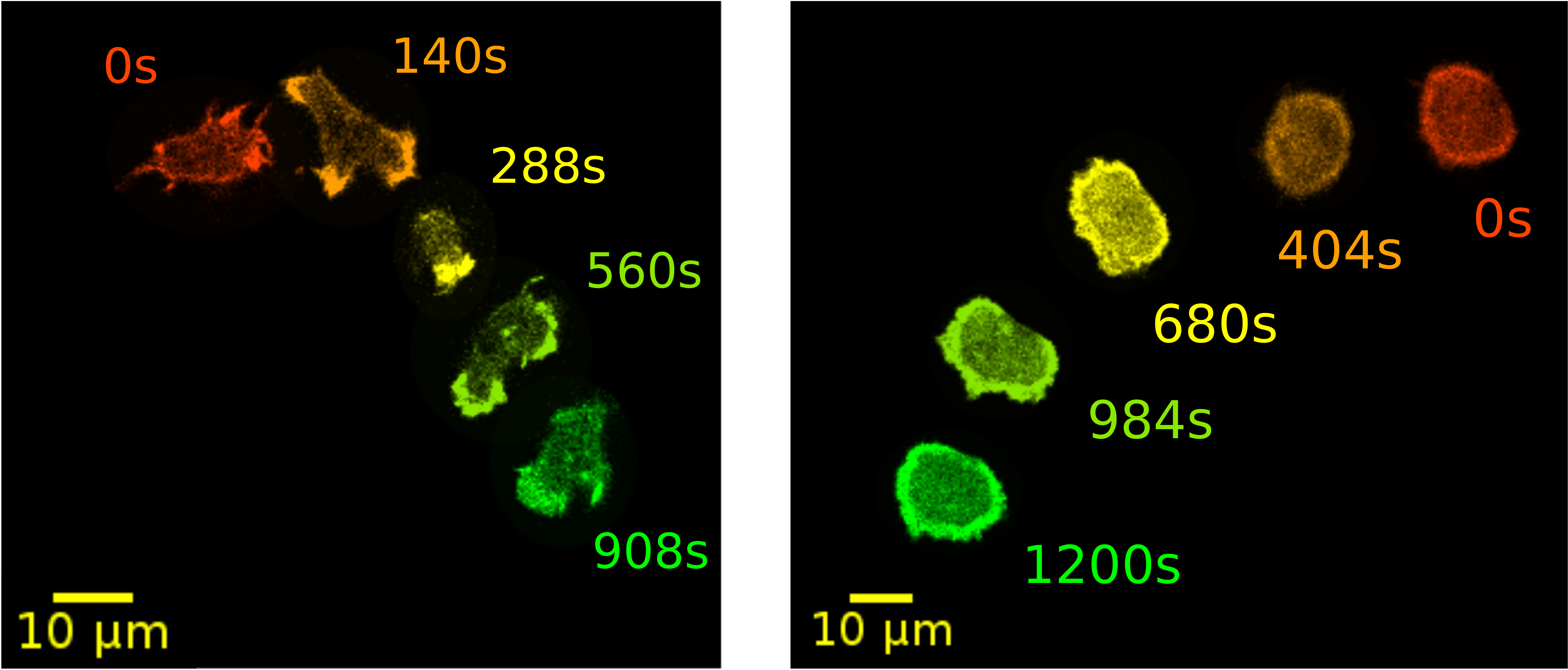}
    \caption{
    Examples of two distinct modes of motility.
    Amoeboid (left) and fan-shaped motion (right) of \dd{} cells expressing Lifeact-GFP as a marker for filamentous actin.
    In each frame, five time points of the same cell track are displayed as an overlay.
    Time is color coded from red, orange and yellow to green.
    \label{fig:1:amoeba_vs_fan}}
\end{figure}

\old{One of the most} \new{A} prominent model organisms to study actin-driven motility is the social amoeba \textit{Dictyostelium discoideum} (\dd{}).
\new{With its genome completely sequenced, \dd{} is readily accessible for a wide range of biochemical and physiological studies.
In addition, many parts of the receptor mediated signaling pathway and the downstream cytoskeletal machinery are conserved between \dd{} and mammalian cells~\citep{artemenkoMovingParadigmCommon2014a,devreotesExcitableSignalTransduction2017}.}
%
It is well known that \dd{} cells can exhibit other types of locomotion besides their common pseudopod-based amoeboid motility.
In particular, they can move in a highly persistent fashion reminiscent of keratocytes.
Here, pseudopodia are largely absent and cells display a stable, kidney-shaped morphology that is elongated perpendicular to their direction of motion.
This so-called fan-shaped motility was first observed in \dd{} cells lacking the \textit{amiB} gene, which is required for aggregation in the course of starvation-induced development~\citep{asano_keratocyte-like_2004}.
In axenic wild-type cells, also the development under low cell density condition may increase the probability of observing fan-shaped cells~\citep{cao_plasticity_2019-1}.
Similarly, in a non-axenic DdB background, an increased proportion of fan-shaped cells can be induced by deletion of the \textit{axeB} gene that encodes the RasGAP NF1~\citep{flemming_how_2020,morenoModelingCellCrawling2020}.
Switches from amoeboid to fan-shaped motility could be even directly induced by decreasing
phosphatidylinositol-4,5-bisphosphate (PtdIns(4,5)P$_2$) levels or by increasing Ras/Rap activity with the help of a rapamycin-induced dimerization system~\citep{miao_altering_2017}.
In Fig.~\ref{fig:1:amoeba_vs_fan}, examples of NF1-deficient DdB cells are displayed, exhibiting amoeboid (left) and fan-shaped motility (right) for comparison. 

In fan-shaped \dd{} cells, a large fraction of the substrate-attached ventral membrane is filled with a composite PtdIns(3,4,5)P$_3$ (PIP$_3$)/actin wave that drives the persistent forward motion~\citep{asano_correlated_2008}.
This was confirmed by the observation of wave-driven cytofission events in oversized \dd{} cells, where it can be clearly seen how the driving wave segment covers the bottom cortex of the emerging daughter cell that moves in a fan-shaped fashion~\citep{flemming_how_2020}.
The fan-shaped mode is thus intrinsically linked to the presence of basal actin waves.
Consequently, treatment with an inhibitor of PI3-kinase (LY294002) that suppresses the formation of basal waves results in a breakdown of fan-shaped motility~\citep{asano_correlated_2008,miao_altering_2017,flemming_how_2020}.
The emergence of actin waves in \dd{} is associated with increased macropinocytic activity, which arises in cells showing hyper Ras activity as a consequence of an NF1 deficiency~\citep{veltman_plasma_2016}.
This is in line with the observation that fan-shaped cells can be induced by synthetically increasing Ras/Rap activity~\citep{miao_altering_2017}.

In previous experiments, where fan-shaped motility was induced by changing developmental conditions or by introducing knockouts, typically only a fraction of the entire cell population adopted the fan-shaped mode~\citep{asano_keratocyte-like_2004,cao_plasticity_2019-1}.
This can be attributed to heterogeneity within the cell population, such as, for example, cell-to-cell variability in gene expression levels, developmental state, or different distributions of other intracellular parameters.
\old{However,} Here we report that switches between amoeboid and fan-shaped modes of migration may also occur spontaneously within the same cell, \new{and we provide a detailed analysis of this reversible switching process.}
\old{without any external trigger such as drug treatment or synthetic modification of intracellular signaling levels by dimerization techniques.}
\new{Specifically,}
when NF1 is knocked out in the non-axenic DdB background, fan-shaped cells are only transiently stable, and cells switch back and forth between fan-shaped and amoeboid modes.
The transition from amoeboid to fan-shaped motility requires the nucleation of an actin wave that grows and eventually fills most of the ventral cortex of the cell to form a stable fan that lives until spontaneous breakdown of the wave occurs.
Not every growing wave successfully locks into a stable fan-shaped configuration, indicating that both the amoeboid and the fan-shaped modes are distinct coexisting states of cortical actin organization.
To account for these observations, we propose an extension of a well-established stochastic reaction-diffusion model of cortical wave dynamics that incorporates bistability of amoeboid and fan-shaped states.

\section{Materials and Methods}

\subsection{Cell culture}
%
For all experiments the non-axenic \dd{} strain DdB NF1 KO was used~\citep{bloomfield_neurofibromin_2015}
\old{The} \new{that carries a} knockout of the \textit{axeB} gene \old{that} encod\new{ing} a homologue of the human RasGAP NF1.
\new{The deficiency in NF1} leads to a phenotype that results in a high percentage of fan-shaped cells~\citep{flemming_how_2020}.
Cells were cultivated in 10~cm dishes or in 125~ml Erlenmeyer flasks with S\o{}rensen's buffer (14.7~mM KH$_2$PO$_4$, 2~mM Na$_2$HPO$_4$, pH~6.0), supplemented with 50~$\mu{}$M MgCl$_2$, 50~$\mu{}$M CaCl$_2$, \old{\textit{Klebsiella aerogenes} at an OD$_{600}$ of 2,} and G418 (5~$\mu${}g/ml) as selection marker for the NF1 KO.
\new{The suspension also contained \textit{Klebsiella aerogenes} at a final OD$_{600}$ of 2, which was achieved by adding a concentrated \textit{Klebsiella aerogenes} solution with an OD$_{600}$ of 20 in a volume corresponding to 1/10 of the final volume of the suspension.
The bacteria were grown in shaking LB medium in a volume of 1 liter to a final OD$_{600}$ of 2, washed 3 times in S\o{}rensen's buffer, and then resuspended in S\o{}rensen's buffer to a concentrated solution with an OD$_{600}$ of 20.}

The \new{DdB NF1 KO} cells were transformed with an episomal plasmid encoding for Lifeact-GFP (either SF99 or SF108) as F-actin marker~\citep{flemming_how_2020}.
The expression vectors are based on a set of vectors for gene expression in non-axenic cells~\citep{paschkeRapidEfficientGenetic2018} \new{and were described in \citet{flemming_how_2020}.}
\new{The plamids were transformed into DdB NF1 KO cells by electroporation as described before (1)}
\old{and were transformed} with an ECM2001 electroporator using three square wave pulses of 500~V for 30~ms in electroporation cuvettes with a gap of 1 mm.
Hygromycin (33~$\mu{}$g/ml) was used as selection marker for the expression plasmid \new{\citep{flemming_how_2020}}.

\subsection{Image acquisition}
In \new{experiments with} the DdB NF1 KO strain, an increased fraction of fan-shaped cells can be observed during the aggregation stage of the \dd{} lifecycle~\citep{morenoModelingCellCrawling2020}.
Aggregation was initiated by removing \textit{Klebsiella aerogenes} to initiate starvation of the \dd{} cells.
DdB NF1 KO cells were grown over night in shaking culture in a volume of 25~ml \new{starting from a density of $3 \times 10^6$ cells/ml to a final concentration of $3 \times 10^7$ cells/ml}.
The next day, remaining bacteria were removed by washing the cells \new{3 times} with S\o{}rensen's buffer \new{(centrifugation at 300 $\times$ g)}.
After washing, the cells were resuspended in 25~ml of S\o{}rensen's buffer and incubated for 3 to 6 hours at 200~rpm in a shaking incubator at 22°~C.
\new{During this incubation time, bacteria that remained after washing were taken up by the \dd{} cells and the developmental stage was initiated.}
Cells were harvested by centrifugation \new{(300 $\times$ g)} and transferred into a 35~mm microscopy dish with glass bottom (FluoroDish, World Precision Instruments) for imaging. The cells were diluted to a density that enabled the imaging of single cell tracks.
A Laser Scanning Microscope (LSM 780, Zeiss, Jena) with a 488~nm argon laser and either a 63$\times$ or 40$\times$ oil objective was used for imaging. 
\new{We observed 48 events, where cells spontaneously switched between amoeboid and fan-shaped modes in a total number of 94 recordings that were each taken over 38 minutes on average.}
\subsection{Data analysis}
\label{sec:data_analysis}
First, discrete sets of points approximating the cell contours were extracted from microscopy images using a modified version of the active contour (snake) algorithm described in~\citep{driscollLocalGlobalMeasures2011, driscollCellShapeDynamics2012}.
An implementation of this segmentation approach can be found in \texttt{AmoePy}, a Python-based toolbox for analyzing and simulating amoeboid cell motility~\citep{schindlerAmoePy2021}.
We then used \texttt{AmoePy} to compute smooth representations of the cell contours and the corresponding kymographs of the contour dynamics, \new{see Fig.~\ref{fig:S1:explanation_kymographs} for an explanation of the kymograph concept}.
The mathematical framework underlying \texttt{AmoePy} was introduced in~\citep{schindler_analysis_2021}.
In short, a Gaussian process regression was used to obtain smooth estimates of the cell contour.
Then, reference points on these contours (so-called virtual markers) were tracked in time relying on a one-parameter family of regularizing flows.
Expansions and contractions were identified based on a novel quantity, the local dispersion.
It measures the local stretching rate of virtual markers on the contour, \new{see Fig.~\ref{fig:S1:explanation_kymographs} for an illustration}.
For this reason, the local dispersion displays a clearly distinct behavior for the two different types of motility: a strong local thinning of markers under the narrow and elongated amoeboid protrusions; and an almost even distribution of markers at the broad leading edge of fan-shaped cells.
These characteristics can be observed, for example, in Fig.~\ref{fig:4:switch_amoeba_fan_exp} panel~C, where the local dispersion kymograph displays large values during amoeboid migration (until $t=1000\,s$), reflected by dark red and dark blue patches.
After switching to the fan-shaped mode, the local dispersion takes small values close to zero, indicated by green patches, confirming that virtual marker thinning is less prominent in this state.

To determine the actin wave area $A_{W}$, we applied an intensity threshold to the fluorescence images, using the  \lq{}Threshold\rq{} function of \texttt{ImageJ}.
Threshold values were chosen based on visual inspection for each image sequence individually.
The wave area was then determined by summing up the pixels exceeding the threshold value.
The fluorescence intensity kymographs in Figs.~\ref{fig:3:failed_switch_exp}D, \ref{fig:S4:minipods_exp}D and \ref{fig:S5:breakdown_wave_exp}D were produced with the \texttt{ImageJ} 'Multi Kymograph' function.

\subsection{Numerical simulations}
\label{sec:model}

\new{The equations of our mathematical model are presented in the Appendix.}
Equations (\ref{pf})--(\ref{delta}) were numerically integrated with periodic boundary conditions using \new{central} finite differences.
The pixel size was defined as $\Delta x=0.15\,\mu$m with an integration step of $\Delta t=0.002$~s.
\new{For the temporal integration of the stochastic partial differential equations we have employed the Euler-Maruyama method. The total area considered for each cell was initialized at $A_0= 113\,\mu$m$^2$ corresponding to a circular cell with radius $r = 6\,\mu$m. 
To produce polarized cells that are immediately motile, initial conditions were chosen such that the activatory component $c$ is asymmetrically distributed across a part of the cell area, see Fig.~\ref{fig:S2:initial_simulation_conditions} in the Supplementary Material.}
The parameters as well as the values and definitions can be found in Table \ref{tab:1}.





\section{Results}
\subsection{Cells spontaneously switch between stable amoeboid and fan-shaped modes of migration}
%
\begin{figure*}
    \centering
        \includegraphics[width=0.7\linewidth]{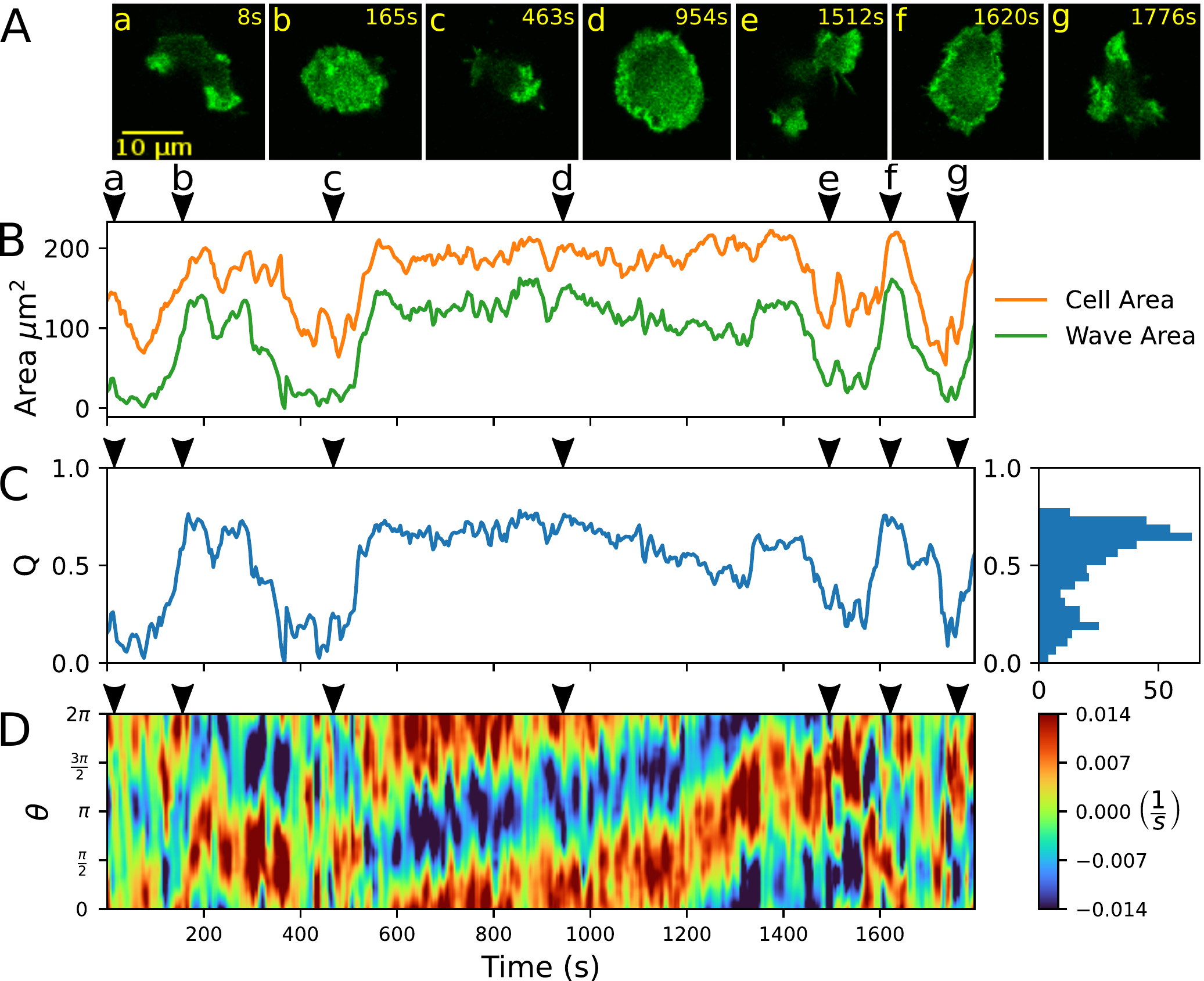}
        \hspace{3mm}
    \caption{
    Spontaneous switching between motility modes.
    (A)~Fluorescence images of a Lifeact-GFP expressing cell, alternating between amoeboid (a, c, e, g) and fan-shaped mode (b, d, f).
    (B)~Time trace of projected cell area $A_{C}$ (orange) and actin wave area $A_{W}$ (green).
    (C)~Time trace of the relative wave area $Q=A_{W}/A_{C}$, where high values are associated with fan-shaped and low values with amoeboid motion.
    The histogram on the right shows the frequency of $Q$-values.
    (D)~Kymograph of the local dispersion along the cell contour.
    Red (blue) indicates extending (contracting) regions along the contour, associated with protruding (retracting) parts.
    Arrowheads above the kymographs mark the time points corresponding to the fluorescence images in~(A).
    \label{fig:2:several_switches_exp}}
\end{figure*}
\begin{figure*}
\centering
    \includegraphics[width=0.7\linewidth]{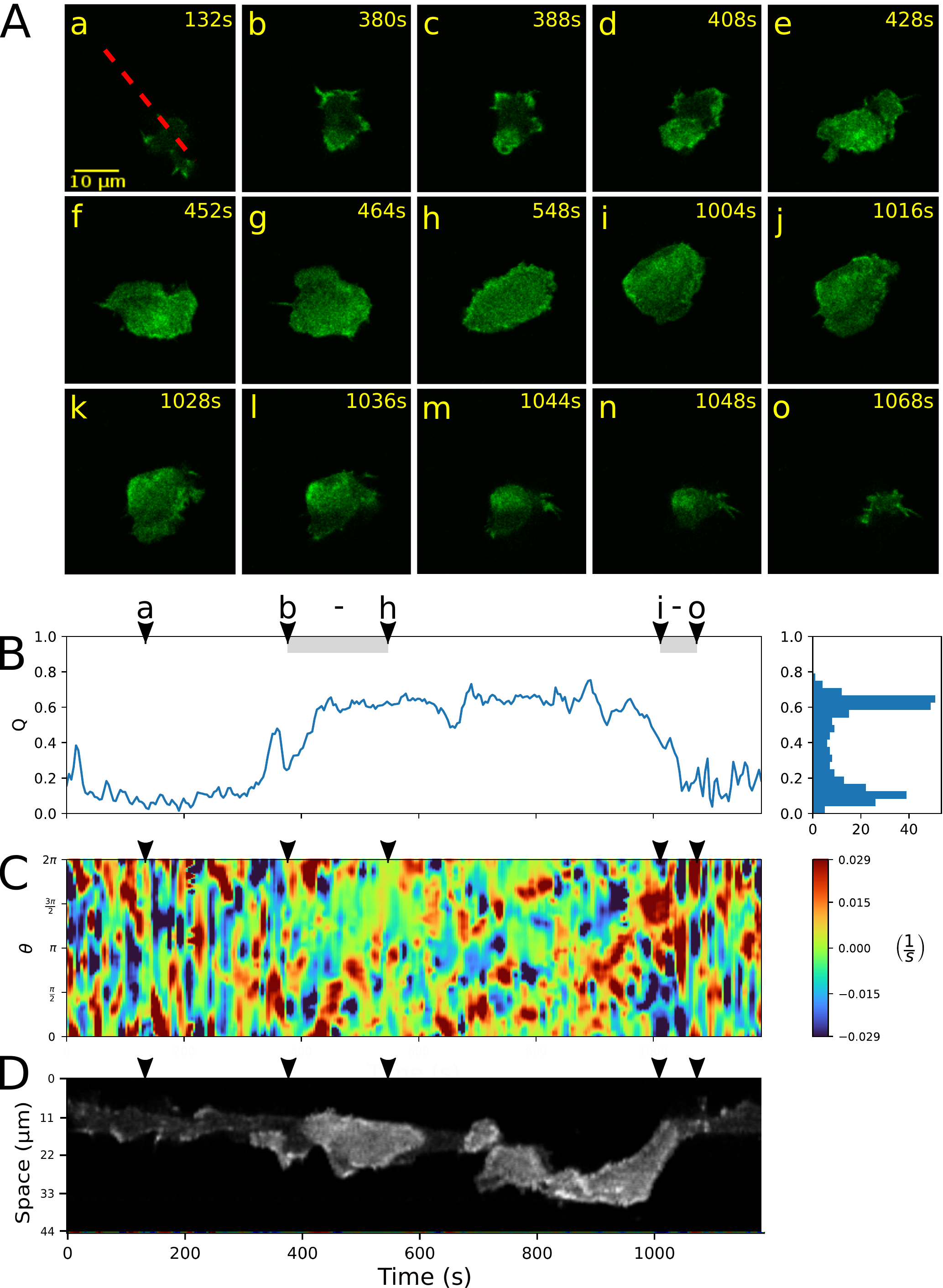}
    \hspace{3mm}
    \caption{
    Growth and decay of actin waves mediate transitions between amoeboid and fan-shaped modes.
    (A)~Starting from a cell in amoeboid mode~(a), nucleation and growth of an actin wave is shown in~(b) to~(h), followed by a rapid wave breakdown in~(i) to~(o).
    (B)~Time evolution of the relative wave area~$Q$.
    The shaded intervals belong to the episodes of actin wave growth~(b-h) and decay~(i-o).
    Histogram on the right-hand side shows the frequency of $Q$-values.
    (C)~Kymograph of the local dispersions along the cell contour.
    (D)~Fluorescence kymograph taken along the red  dashed line shown in the first panel of~(A).
    Arrowheads above the kymographs indicate the time points of the corresponding fluorescence images in~(A).
    \label{fig:3:failed_switch_exp}}
\end{figure*}

Previously, \old{switches between amoeboid and a} \new{stable} fan-shaped \old{modes of migration} \new{cells} have been observed as a consequence of \new{genetic modifications, specific developmental conditions,} \old{drug treatment} or synthetic changes in the localization of intracellular signaling components~\citep{asano_correlated_2008,miao_altering_2017,cao_plasticity_2019-1}.
Here, we report that \dd{} cells may also show spontaneous, \new{reversible} switches between amoeboid and fan-shaped motility \old{in the absence of any external trigger}.
Our observations were recorded in non-axenic DdB cells, where the \textit{axeB} gene was disrupted, so that the RasGAP NF1 could not be expressed and, consequently, Ras activity was increased~\citep{veltman_plasma_2016}.
The spontaneous switches occurred in the developed state after two to six hours of starvation.
%
In Fig.~\ref{fig:2:several_switches_exp}, an example is displayed.
The sequence of fluorescence images in Fig.~\ref{fig:2:several_switches_exp}(A) shows a cell undergoing a series of several consecutive switching events.
    
Besides manual classification based on visual inspection by an observer, switches between amoeboid and fan-shaped modes can be identified by changes in the projected cell area $A_{\mathrm{C}}$~\citep{miao_altering_2017}.
In Fig.~\ref{fig:2:several_switches_exp}(B), the area $A_{\mathrm{C}}$ is shown over time (orange line).
In the fan-shaped state, the cell adopts a flattened, \new{spread-out} geometry \old{and spreads across the substrate, due to the protruding forces exerted by the driving wave segment onto the cell contour}.
This \old{results in} \new{leads to} high values of $A_{\mathrm{C}}$ for the fan-shaped state, in contrast to lower values, \old{associated with} \new{resulting from} the more compact geometry during amoeboid migration. 
As the fan-shaped state is associated with an actin wave that spreads across the ventral cortex, we also determined the cortical area $A_{\mathrm{W}}$ that is covered by the actin wave at each time point of our recordings, see the green line in Fig.~\ref{fig:2:several_switches_exp}(B).
This was accomplished by expressing a fluorescent marker for filamentous actin (Lifeact-GFP), \old{so that} \new{thus} waves could be identified based on increased intensity values in the fluorescence images.
Qualitatively, the temporal evolution of the wave area $A_{\mathrm{W}}$ follows the same trend as the \old{overall} \new{projected} cell area $A_{\mathrm{C}}$.
As the absolute values of cell and wave areas may differ from cell to cell, we introduced the dimensionless ratio $Q = A_{\mathrm{W}}/A_{\mathrm{C}}$, which we call the relative wave area, to distinguish the two migration modes.
In Fig.~\ref{fig:2:several_switches_exp}(C), the time evolution of the relative wave area $Q$ is shown.
The $Q$-values are confined between 0 and 1 because the wave area $A_{\mathrm{W}}$ cannot exceed the projected cell area $A_{\mathrm{C}}$.
Typically, values of $Q < 0.5$ are associated with the amoeboid mode of motion, whereas values of $Q > 0.5$ indicate a fan-shaped state.
The two states are reflected by two peaks in the frequency histogram of $Q$-values that is displayed on the right-hand side of Fig.~\ref{fig:2:several_switches_exp}(C).
\new{This classification into amoeboid and fan-shaped cells based on the relative wave area $Q$ is not specific to our data set but can be generally applied also when these migratory modes are induced by other conditions.
To demonstrate this, we have calculated $Q$-values from previously published data by others, where amoeboid and fan-shaped modes were induced separately by synthetic clamping of the intracellular PtdIns(4,5)P$_2$ levels~\citep{miao_altering_2017,ghabache_coupling_2021}.
The Q-values of amoeboid and fan-shaped cells from these data sets are in good agreement with the $Q$-values that were observed in our own experiments, see Fig.~\ref{fig:S3:comparison_of_Q_values} in the Supplementary Material.}

The kymograph in Fig.~\ref{fig:2:several_switches_exp}(D) displays the local dispersion computed for 400 virtual markers placed along the cell circumference that was parameterized by an angular coordinate ranging from 0 to $2\pi$.
The dispersion is a measure of the local stretching rate of the contour and provides a sensitive measure to identify protruding and retracting regions of the cell border~\citep{schindler_analysis_2021}.
In particular, positive (negative) values of the local dispersion correspond the protruding (retracting) parts of the contour.
For cells that move in a fan-shaped fashion, the leading and trailing edges are reflected by broad regions in the kymograph that show persistently protruding and retracting activity, respectively, see for example the fan-shaped episode between 600 and 1300~s in Fig.~\ref{fig:2:several_switches_exp}(D).
In contrast, amoeboid movement results in a random distribution of smaller patches of activity along the contour that can be associated with extending and retracting pseudopodia.
\new{Note that also at the leading edge of fan-shaped cells, occasionally small short-lived protrusions emerge.
However, they typically do not interfere with the overall stable movement of the fan-shaped cell, see Fig.~\ref{fig:S4:minipods_exp} in the Supplementary Material for an example.}

\begin{figure*}
\centering
    \includegraphics[scale=0.55]{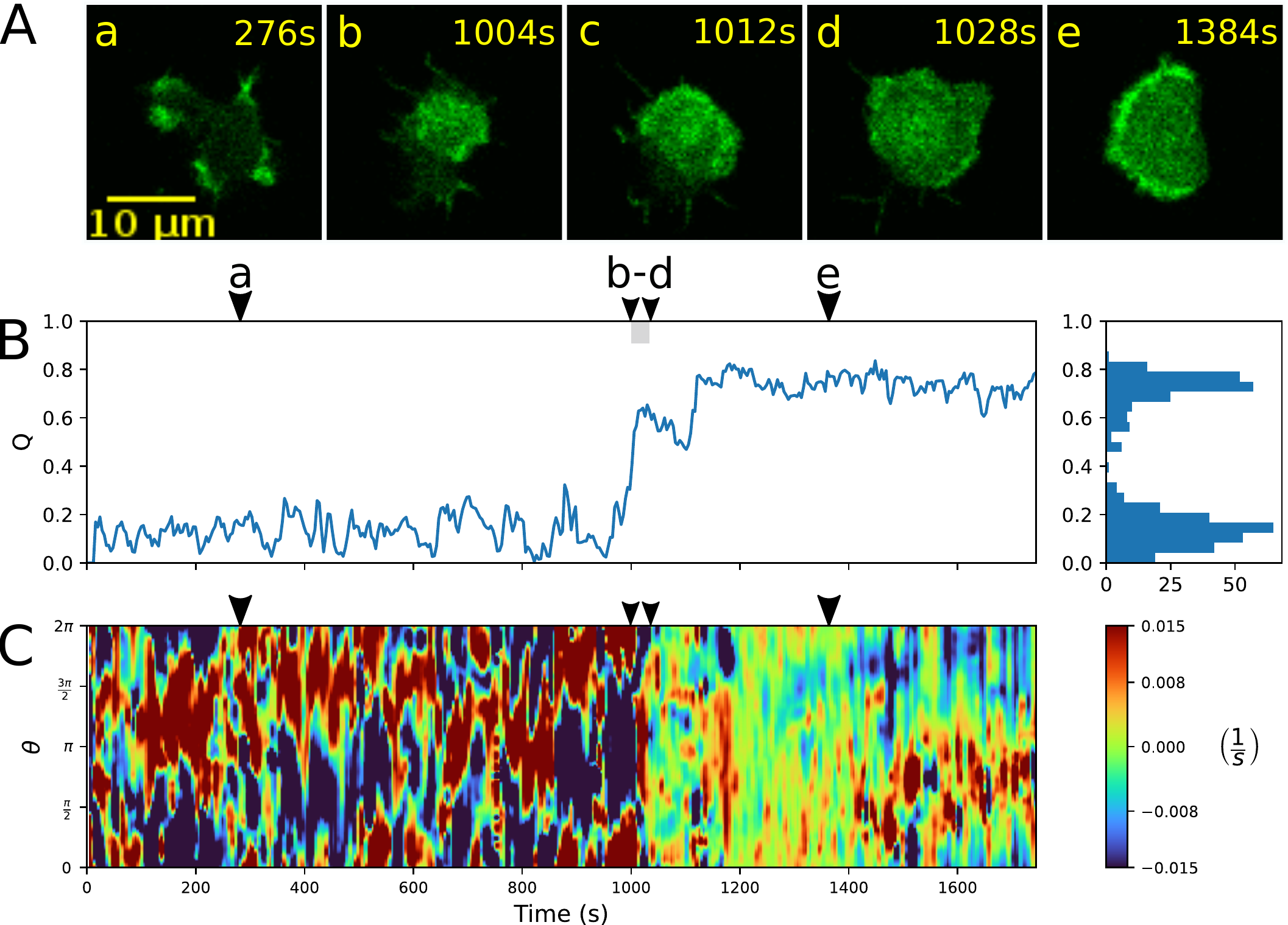}
    \hspace{3mm}
    \caption{
    Transition from amoeboid to fan-shaped mode.
    (A)~Sequence of fluorescence images starting in amoeboid mode~(a) and switching from~(b) to~(d) into the fan-shaped mode~(e).
    (B)~Time evolution of the relative wave area~$Q$.
    Histogram on the right-hand side shows the frequency of $Q$-values.
    (C)~Local dispersion kymograph taken along the cell contour.
    Arrowheads above the kymographs indicate the time points of the corresponding fluorescence images in~(A).
    \label{fig:4:switch_amoeba_fan_exp}}
\end{figure*}
\subsection{Fan-shaped cells are formed by wave nucleation and growth from amoeboid cells}
We further focus on a more detailed characterization of the switching events between amoeboid and fan-shaped modes of locomotion.
Typically, the bottom cortex of DdB wild type cells exhibits a dynamically fluctuating F-actin density with long-lived actin foci and
occasional bursts of larger actin patches that transiently emerge and decay~\citep{flemming_how_2020}.
In contrast to the wild type, in NF1 KO cells that show hyper Ras activity, these actin patches may increase in size and become the nucleus of a growing ring-shaped actin wave.
In Fig.~\ref{fig:3:failed_switch_exp}A, an example of such an event is shown.
Starting from the amoeboid state (a), a wave is nucleated close to the lower border of the cell (b), grows in size (c,d), and spreads across the ventral cortex until the entire projected area of the cell is filled by the actin wave~(h).
During this growth process, the actin wave pushes the cell border outwards, eliminating smaller pseudopodial protrusions and irregularities of the cell shape, until it converges to the roundish, spread-out morphology of a fan-shaped cell.
This is reflected by an increase in the relative wave area $Q$ from frame (b) to (h) in Fig.~\ref{fig:3:failed_switch_exp}B.
During the same time interval, the kymograph of the local dispersion in C shows that the numerous irregular protrusions that can be observed during amoeboid migration, are increasingly suppressed, giving rise to a smoother cell contour with only a few broader protrusions.
A fluorescence kymograph taken along the dashed red line in (a) confirms that this time interval coincides with a growing wave, as can be seen by the increased intensity levels of the Lifeact-GFP label inside the wave area, see Fig.~\ref{fig:3:failed_switch_exp}D.

Once a fan-shaped geometry has been reached, we observed two different scenarios for the further evolution.
On the one hand, the extended wave can be unstable, giving rise to breakup of the wave area and an eventual decay back to the amoeboid state.
During this process, the relative wave area $Q$ fluctuates around an elevated level, see Fig.~\ref{fig:3:failed_switch_exp}B between (h) and (i), accompanied by an irregular dynamics of the cell contour displayed in C.
The fluorescence kymograph in panel D confirms that the wave area is unstable during this time interval.
The final return to the amoeboid state is associated with a complete decay of the wave between time frames (i) and~(o) and can be identified by a steep decrease in the relative wave area $Q$ in panel~B.

Alternatively, once the wave fills the entire bottom cortex, the cell can \say{lock} into a stable fan-shaped state, characterized by a large, extended wave segment that maintains a roundish, elongated morphology and pushes the cell forward in a persistent fashion.
Details of the transition to a stable fan can be \old{inferred from} \new{seen in} Fig.~\ref{fig:4:switch_amoeba_fan_exp}.
Starting from an amoeboid state (a), a rapid transition to the fan-shaped mode induced by a growing wave is captured in time frames (b) to~(d), which is accompanied by an increase in the relative wave area $Q$ and a smoothing of the cell contour, see Figs.~\ref{fig:4:switch_amoeba_fan_exp}B and~C.
In this case, the fan initially remains stationary and resumes movement only after 1400~s, as can be seen from the formation of a persistent leading edge in the local dispersion kymograph \new{in panel~C} (red patches steadily located between $0$ and $\pi$ on the contour). \old{in panel~C}.
%
%
The transition from fan-shaped back to amoeboid motility is \new{typically} initiated by a \old{rapid} \new{spontaneous} breakdown of the basal actin wave, see supplementary Fig.~\ref{fig:S5:breakdown_wave_exp} for an example.
It proceeds in a similar manner as the collapse of an unstable fan shown in Fig.~\ref{fig:3:failed_switch_exp}.
%
\old{Most commonly, breakdown of the actin wave occurred spontaneously.}

\subsection*{\old{Small protrusions emerge at the leading edge of fan-shaped cells}}
\label{sec:minipods_old}
%

\old{
At the leading edge of persistently moving fan-shaped cells, we repeatedly observed the emergence of small, short-lived protrusions.
To distinguish them from the pseudopodia of amoeboid migration, we hereafter call them minipods.
In Fig.~\ref{fig:S4:minipods_exp}A, the formation and decay of a minipod is illustrated in a sequence of snapshots (a)--(e) from the fluorescence recording of a stable fan-shaped cell.
The entire event extends over a short time window of about 36~s, which is typical for the life-time of these structures.
Due to their small size, the formation of minipods does not affect the overall value of the relative wave area $Q$, see for example Fig.~\ref{fig:S4:minipods_exp}B, where $Q$ remains around $0.6$ for the entire measurement time.
However, in the local dispersion kymograph of the cell contour in panel C, minipod formation can be clearly detected as a strongly localized, protruding region (red), with a simultaneous increased retraction along the rest of the contour (blue). See the time period between (a) and~(e), as well as the time period shortly after $t=200$~s.
}

\old{
The growing minipod is accompanied by a decrease of the fluorescence intensity in its vicinity.
In particular, the front of the actin wave that pushes the cell forward is directly affected by the growing minipod, such that the leading wave segment is disrupted, indicating a depletion of F-actin, which causes a breakup of the wave segment at the position of minipod formation, see Fig.~\ref{fig:S4:minipods_exp}A at time frames (b)--(d).
We thus conclude that the growing minipod competes with the wave for the common pool of actin, consuming some of it in its surrounding in order to push the plasma membrane outwards. 
When the minipod has decayed, the disrupted actin segment at the wave front heals and recovers its initial fluorescence intensity.
This can be also seen in the fluorescence intensity kymograph in Fig.~\ref{fig:S4:minipods_exp}D, taken along the red dashed line displayed in the first snapshot of panel~A.
}

\old{
The kymograph in Fig.~\ref{fig:S4:minipods_exp}D also shows that the decaying minipod is actively retracted.
It protrudes from the leading edge with a speed that is faster than the propagation velocity of the cell, then stops growing, and finally is retracted to join the advancing front of the fan that is driven by the actin wave.
Whether minipods actively contribute to the locomotion of the fan or simply decorate the leading edge without any impact on the net propagation of the cell cannot be decided on the basis of the available data.
}




\begin{figure*}
\centering
    \includegraphics[scale=0.55]{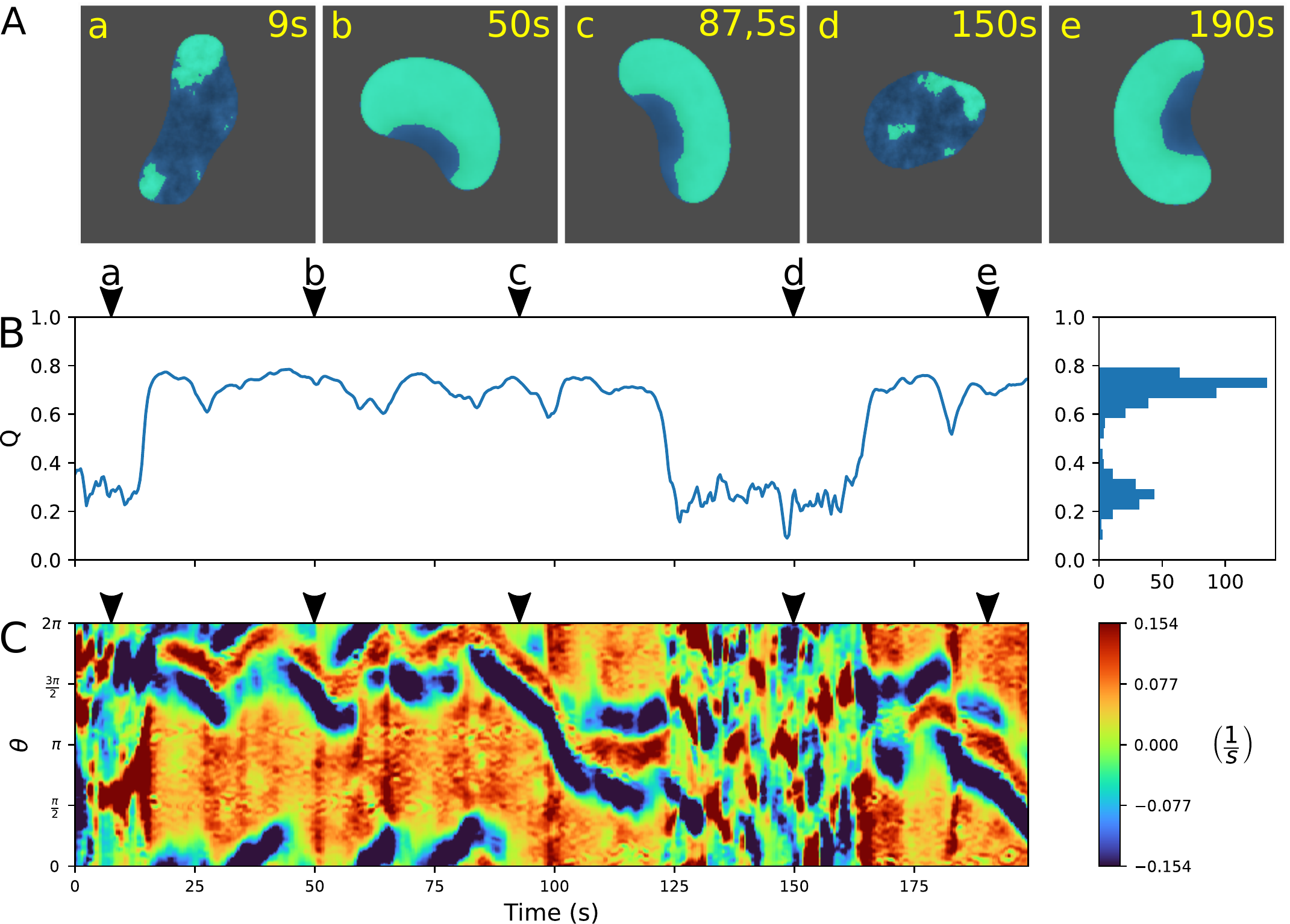}
    \hspace{3mm}
    \caption{Predicting spontaneous switching between motility modes by the model.
    (A)~Snapshots of a cell alternating between amoeboid (a, d) and fan-shaped mode (b, c, e).
    (B)~Time trace of the relative wave area $Q=A_{W}/A_{C}$, where high values are associated with fan-shaped and low values with amoeboid motion.
    Histogram on the right shows the frequency of $Q$-values.
    (C)~Kymograph of the local dispersion taken along the cell contour.
    Arrowheads above the kymographs mark the time points corresponding to the fluorescence images in~(A).
    \new{The concentration constraint parameter was set to $M=0.0015$ and the threshold value to $Q_{Th}=0.45$;
    for all other model parameters see Table~\ref{tab:1}.}
    \label{fig:5:several_switches_sim}}
\end{figure*}

\subsection{Numerical simulations of a phenomenological motility model capture the experimental findings}
We performed numerical simulations using a well-established phenomenological model of actin-driven cell motility~\citep{alonso_modeling_2018}.
The model is based on a noisy bistable reaction-diffusion system for the intracellular dynamics that is combined with a dynamic phase field to take the evolution of the cell shape into account, see \old{Sec.~\ref{sec:model}} \new{the Appendix} for \old{the detailed} \new{a presentation of the} model equations.
The patterns of cortical activity are modeled by a single dynamical species $c$ that can be associated with activatory components of the signaling pathway, such as activated Ras, PI3K, and PIP$_3$.
Thus, the component $c$ indirectly reflects also the cortical F-actin patterns that drive the cell contour dynamics, which is achieved in the model by a coupling of the local concentration of $c$ to the dynamics of the confining phase field.

We previously found that this model shows both amoeboid and fan-shaped motility modes depending on the choice of model parameters, such as the noise intensity and the average cortical coverage with the activatory component $c$~\citep{morenoModelingCellCrawling2020}.
This and previous findings by others \old{seem to indicate} \new{suggest} that within a cell population, different motility modes may arise due to cell-to-cell variability \old{see also Sec.~\ref{sec:introduction}}~\citep{asano_keratocyte-like_2004,cao_plasticity_2019-1}.
However, considering our experimental observation \old{of} \new{that individual} cells frequently switch back and forth between both modes of locomotion, we assume that these modes \new{are} coexisting \old{as stable} behavioral \old{states} \new{traits} within the same cell, so that noise induced transitions may occur between them. 

In order to take these findings into account, we extended our previously established model.
Specifically, the average cortical coverage with the activatory species $c$, $C_0 = Q_0 \int\phi\,dA$, that was a parameter in the previous version of the model, can now dynamically change between two stable states of low and high $c$-coverage, respectively, see Eq.~(\ref{eq:cases}).
The low coverage state corresponds to the amoeboid mode, where the cortical activity is dominated by small, short-lived bursts of actin activity, whereas the high coverage state is associated with the fan-shaped mode, where most of the ventral cortex is covered by a large, stable actin wave.

In numerical simulations of the extended model, we indeed observed rapid, spontaneous switches between amoeboid and fan-shaped locomotion.
An example is displayed in Fig.~\ref{fig:5:several_switches_sim}, where the individual snapshots shown in panel A can be assigned to either the amoeboid mode that shows small irregular patches of cortical activity (a and d), or to the fan-shaped mode that is dominated by a large wave segment, imposing a stable, elongated cell shape (b, c, and~e).
Similar to the experimental observations, also in simulations the two states can be distinguished based on the value of the relative wave area $Q$.
In Fig.~\ref{fig:5:several_switches_sim}B, the time evolution of $Q$ is displayed, clearly highlighting the transitions between the two stable states.
They are characterized by their respective stable $Q$-values.
Note that, in the numerical simulations, the $Q$-values of the amoeboid and fan-shaped states correspond to the average cortical coverages $Q_0$ in the respective states.
They are model parameters chosen in agreement with the experimental observations, see Eq.~(\ref{eq:cases}).

\old{We also computed kymographs of the local dispersion} 
To further characterize the dynamics of the simulated cell contours, \new{we computed kymographs of the local dispersion.}
Similar to the experiments, the contour evolution is dominated by small irregular protrusions in the amoeboid state that is distinct from the regular pattern with stable leading and trailing edges in the fan-shaped state, see Fig.~\ref{fig:5:several_switches_sim}C.
\old{We note, however, that at the level of the detailed contour dynamics, the agreement between the experimental and numerical results remains qualitative.
In particular, many of the fine-scale structures, such as the minipods described in Sec.~\ref{sec:minipods}, cannot be reproduced with the current extended model.}
Additional numerical examples of switching events between amoeboid and fan-shaped locomotion can be seen in the Supplementary Information, where Fig.~\ref{fig:S6:switch_amoeba_fan_sim} shows a switch from amoeboid to fan-shaped mode and Fig.~\ref{fig:S7:breakdown_wave_sim} a transition from a fan to an amoeba, induced by spontaneous breakdown of the driving wave segment.
\new{To further asses the performance of our model, we analyzed the shape of amoeboid and fan-shaped cells by measuring the distance from each point on the cell border to the center of mass of the cell and compared the time evolution of this quantity for experimentally measured and numerically simulated cells.
The resulting kymographs and profiles are displayed in Figs.~\ref{fig:S9:distance_to_center_exp} and~\ref{fig:S10:distance_to_center_sim}.
They show that the shape evolution of both the rapidly changing, irregular amoeboid cells as well as the stable elongated fan-shaped cells is closely reproduced by our model simulations.}

\begin{figure}[t] 
\begin{center}
    \includegraphics[width=0.7\linewidth]{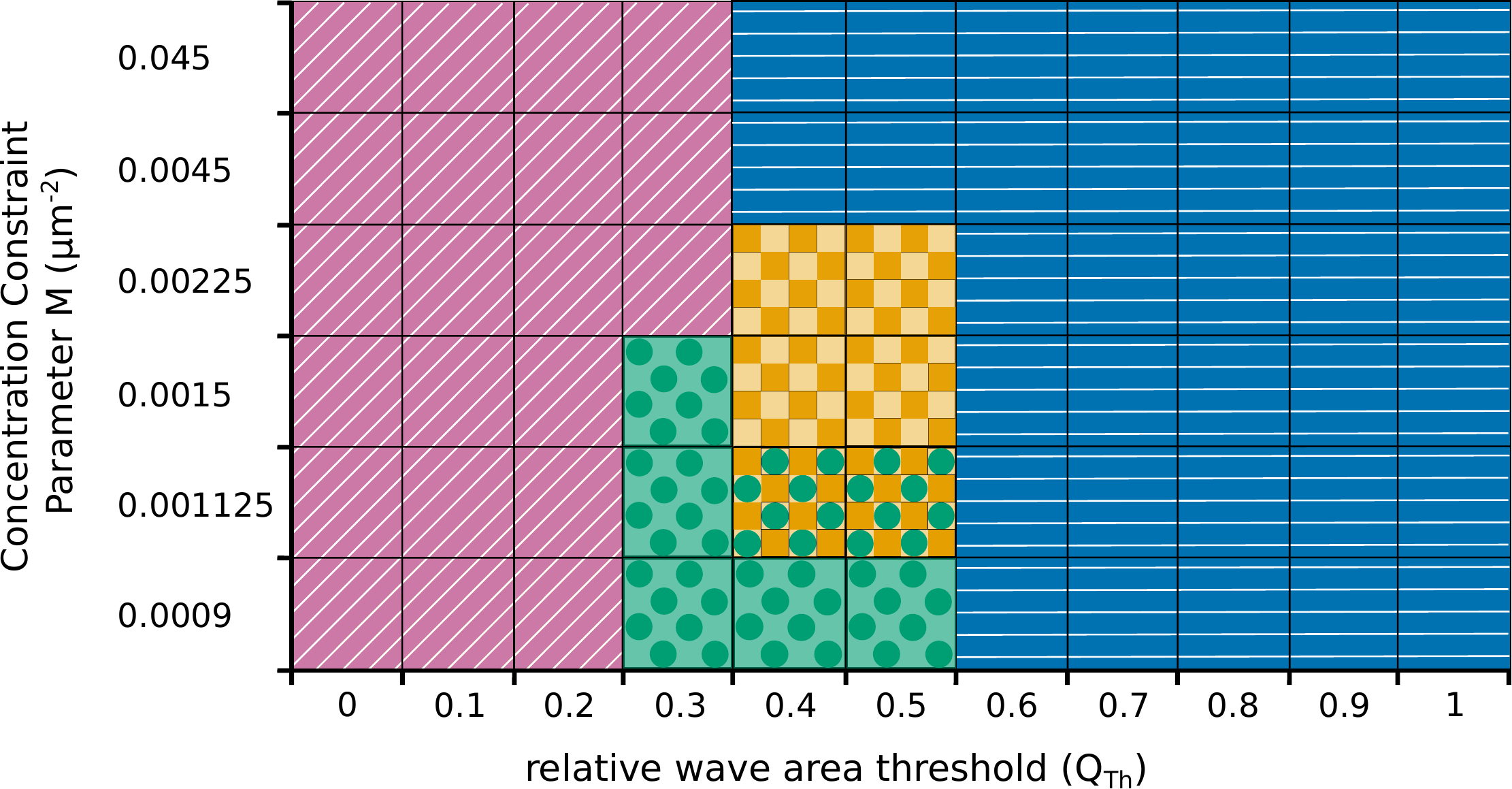}
\end{center}
\caption{Parameter plane spanned by the threshold value $Q_{Th}$ in the relative wave area and the concentration constraint parameter $M$ that controls regulation of the cortical coverage with the activatory component $c$.
Simulations result either in fan-shaped cells (purple oblique lines), in amoeboid cell (blue horizontal lines), or in stationary circular cells (green circles).
Only at intermediate parameter values, spontaneous switching between amoeboid and fan-shaped cells is observed (yellow squares). In some cases, two behaviours coexist and the particular dynamics depends on the initial condition (yellow squares and green circles).
\label{fig:6:phase_diagram}}
\end{figure}

We also systematically explored the impact of changes in the model parameters on our findings, see Fig.~\ref{fig:6:phase_diagram} for an illustrative summary.
As expected, our numerical simulations revealed that the switching behavior critically depends on the choice of the threshold value $Q_{Th}$. For low values of $Q_{Th}$ only fan-shaped cells are observed (marked by purple oblique lines in Fig.~\ref{fig:6:phase_diagram}) and for large values cells exclusively move in an amoeboid fashion (marked by blue horizontal lines).
Repeated switching between both states only occurs at intermediated $Q_{Th}$-values around 0.5 (yellow squares).
However, whether the switching regime at intermediate $Q_{Th}$-values is observed or not also depends on the value of the concentration constraint parameter $M$ that determines how tightly the cortical $c$-coverage is regulated to match the target value of the amoeboid or fan-shaped state.
As can be seen in Fig.~\ref{fig:6:phase_diagram}, a tight regulation suppresses the switching regime.
\new{In this case, the cell remains locked to the amoeboid or the fan-shaped state}, so that for intermediate $Q_{Th}$-values a sharp boundary between permanently amoeboid and permanently fan-shaped motility is observed.
Only for \old{decreasing} values of $M$ below $0.0045$, the switching regime emerges (yellow squares).
It is, however, also bounded towards lower values of $M$ and disappears if $M$ is reduced \new{to values below 0.0015}.
\old{by a factor of 50 or more from the upper boundary at $0.045$.}
At the same time, a new dynamical regime emerges in this parameter range, where the cell converges to a stationary circular shape, completely filled with a $c$-rich domain (marked by green circles).

%

\section{Discussion}
Our experimental recordings show that motile \textit{D.~discoideum} cells with a deficiency in the RasGAP NF1 spontaneously switch back and forth between amoeboid and fan-shaped modes of motion.
This observation suggests that amoeboid and fan-shaped modes can be seen as \old{stable} coexisting behavioral \old{states} \new{traits} of the same cell rather than signatures of heterogeneity in a cell population, where different modes of motility occur in different cells due to cell-to-cell variability.
Thus, repeated noise-induced switches between these two \old{stable} states can spontaneously occur in the same cell and do not require changes in the intracellular parameters by drug treatment or altered gene expression to initiate a transition.

This was less obvious from previous experimental observations of the fan-shaped mode, where typically fractions of fan-shaped cells across populations were reported.
Examples include \textit{amiB}-null cells~\citep{asano_keratocyte-like_2004,asano_correlated_2008}, axenic wildtype cells developed under low density conditions~\citep{cao_plasticity_2019-1}, and, more recently, also knockout cells deficient in the Arp2/3 inactivating factor GmfA (glia maturation factor)~\citep{fujimotoDeletionGmfAInduces2022}.
However, already in these earlier recordings, occasional switches at the single cell level can be seen~\citep{cao_plasticity_2019-1,fujimotoDeletionGmfAInduces2022}.
We assume that in these cases, the fan-shaped mode is more stable, so that switches within the same cell are less likely to be observed.
Similarly, externally induced parameter changes, such as synthetically \old{increased} \new{clamped} levels of phosphatidylinositol-4,5-bisphosphate (PtdIns(4,5)P2) and Ras/Rap activity~\citep{miao_altering_2017}, or a reduced protrusive strength of the actin cortex by treatment with latrunculin~B~\citep{cao_plasticity_2019-1},
may result in altered life times of the fan-shaped state.

Recently, the different migratory modes have also been characterized in terms of the traction force patterns that are associated with their intracellular actomyosin distributions, revealing that even within the population of fan-shaped cells, different propulsion mechanisms can be distinguished~\citep{ghabache_coupling_2021}.
Note also that in several of these earlier studies, a third so-called oscillatory mode was reported, where cells adopted a periodically breathing, circular, and almost non-motile configuration~\citep{miao_altering_2017,cao_plasticity_2019-1,ghabache_coupling_2021}.
Longer periods of this behavior were not observed in our recordings but can be most likely associated with the stationary circular morphologies observed in our model simulations, see Fig.~\ref{fig:6:phase_diagram} (green circles).
In our experiments, stationary circular cells that resemble the oscillatory mode emerged only occasionally and transiently as part of a switching event.

Periodic dynamics reminiscent of wave-mediated switching between amoeboid and fan-shaped modes has been observed already earlier in axenic cells~\citep{gerisch_different_2011,gerisch_pip3_2012}.
Here, time intervals without any prominent actin wave activity alternate with transient periods of wave growth and expansion across the entire ventral cortex.
In the light of our findings, these episodes can be interpreted as incomplete transitions from amoeboid to fan-shaped mode, where a growing wave eventually occupies the entire bottom cortex but does not succeed in establishing a stable fan.
This is in line with the observation that the fan-shaped mode is only rarely seen in axenic wildtype cells under normal culture conditions.

\old{
Our initial conjecture of two stable, coexisting modes of locomotion is further supported by the observation of minipods described in Sec.~\ref{sec:minipods}.
They have been briefly mentioned earlier \mbox{\citep{flemming_how_2020}} and can be also induced by chemotactic cues~\mbox{\citep{ecke_co-existence_2017}}.
These small protruding structures that randomly emerge at the leading edge of fan-shaped cells, resemble remnants of pseudopodia, possibly indicating that the pseudopod generator of amoeboid motility coexists with the wave-driven fan-shaped mode of locomotion.
In addition, our fluorescence recordings suggest that both structures, the actin wave at the leading edge of a fan and the minipods, compete for the common pool of actin, such that a growing minipod may disrupt a neighboring wave segment by actin depletion, see Fig.~\ref{fig:S4:minipods_exp}.
To what extent such minipod-induced perturbations of the driving wave segment may contribute to a breakdown of the actin wave and a subsequent transition from fan-shaped to amoeboid motility cannot be decided from our current data but seems a plausible scenario.
}



To model the motility of adherent cells, nonlinear reaction-diffusion systems are a widely used choice.
They successfully capture the dynamics of cell polarity~\citep{jilkine_comparison_2011}, which is often related to conservation of some of the participating components~\citep{moriWavePinningCellPolarity2008,otsujiMassConservedReaction2007,bernitt_fronts_2017,halatek_self-organization_2018}, to mutual inhibition~\citep{matsuoka_mutual_2018}, or to external chemical gradients~\citep{beta_bistable_2008,iglesias_navigating_2008}.
In combination with an auxiliary phase field to account for the moving cell boundary, this method was first applied to keratocyte motility~\citep{shao_computational_2010,ziebert_model_2012,shao_coupling_2012,camley_crawling_2017} and later extended to neutrophil and {\it Dictyostelium} morphodynamics~\citep{najem_phase-field_2013,moure_computational_2016,alonso_modeling_2018,imotoComparativeMappingCrawlingcell2021}.
Here, dynamic wave patterns in the intracellular reaction-diffusion system drive the formation of membrane protrusions to propagate the cell forward.
Also, the interactions among adjacent cells and solid boundaries were incorporated into this modeling framework~\citep{lober_collisions_2015,kulawiak_modeling_2016,morenoSingleCollectiveMotion2022}, as well as extensions to three dimensions~\citep{cao_minimal_2019,winkler_confinement_2019}, corresponding to locomotion in enclosed environments~\citep{nagel_geometry-driven_2014}.

Several models were proposed that exhibit both amoeboid and fan-shaped modes, including a lattice model at the level of the actin dynamics~\citep{nishimura_cortical_2009}, a level set approach relying on a biased excitable signaling network~\citep{miao_altering_2017}, as well as phase field models that focus on nonlinear signaling kinetics~\citep{morenoModelingCellCrawling2020} or incorporate mechanochemical coupling and force generation~\citep{cao_plasticity_2019-1,ghabache_coupling_2021}.
To account for our experimental observations, we extended an established motility model by incorporating the coexistence of two stable modes of locomotion (bistability).
We chose a phenomenological model that \old{we recently} \new{was} designed to explain the cell-to-cell variability in the motion patterns of motile {\it D.~discoideum} cells~\citep{alonso_modeling_2018} and that was later found to exhibit both amoeboid and fan-shaped modes of locomotion~\citep{morenoModelingCellCrawling2020}.
This model is based on a noisy bistable reaction-diffusion system to mimic the intracellular dynamics, combined with a dynamic phase field to take the cell shape dynamics into account.
It has also been extended to a two-variable activator-inhibitor version to describe wave-driven cytofission events~\citep{flemming_how_2020}.

In the fan-shaped state, most of the substrate-attached bottom membrane is covered by a composite PIP$_3$/actin wave, whereas in the amoeboid state, only smaller patches of activity are observed along the cell border.
Consequently, we included an additional bistability into the model that allowed for a noise-induced switching between an amoeboid state with low coverage of the activatory component $c$, and a fan-shaped state, where a large fraction of the cell area is covered with $c$.
In this way, we could qualitatively reproduce our experimental findings.
Furthermore, our modeling study predicted that spontaneous switching between the two modes is only observed
\old{at intermediate efficiencies of the regulatory pathway that controls the size of the basal PIP$_3$/actin patches, represented by the concentration constraint parameter $M$ in the model.}
\new{in a limited window of intermediate values of the model parameters $Q_{Th}$ and $M$, see Fig.~\ref{fig:6:phase_diagram}.
Even though these parameters cannot be unambiguously linked to experimentally accessible quantities, we may rationalize their meaning in the light of the current state of our knowledge.
Fan-shaped cells occur under conditions, where the concentration of typical cell front markers, such as PIP$_3$ or active Ras, is upregulated, e.g., by synthetically clamping the intracellular PtdIns(4,5)P2 at low or the Ras/Rap activity at high levels~\citep{miao_altering_2017}.
In our model, the threshold parameter $Q_{Th}$ sets the probability that the cell will adopt such a state of high front marker levels:
the lower the value of $Q_{Th}$ the higher the probability that the cell will converge to a fan-shaped state.
For this reason, regular switching between amoeboid and fan-shaped modes can only occur at intermediate values of $Q_{Th}$.
Altering the value of $Q_{Th}$ in the model thus corresponds to an experiment, where the probability to switch to the fan-shaped mode is gradually changed.}
This can be potentially tested in future dosing experiments that monitor the switching rate as a function of intracellular PIP$_3$ or Ras levels\old{, which can be synthetically controlled by rapamycin-induced dimerization reactions}
\new{once a technique becomes available that allows for a quantitative tuning of such intracellular signaling activities.}

\new{
The concentration constraint parameter $M$, on the other hand, controls how tightly the cell regulates the intracellular signaling levels to the target values that correspond to the amoeboid or fan-shaped states.
For large values of $M$ there is little chance to escape from an established amoeboid or fan-shaped state, so that only for sufficiently low values of $M$ switches may occur.
It remains unclear how this parameter can be directly changed in experiments.
But the characteristic time scale of this regulatory process that is controlled by the parameter $M$ could be assessed by measuring the response time of basal actin waves to changes in intracellular PIP$_3$ or Ras levels.
Again, a quantitative dosing technique, which is currently not available in sufficient precision, would be required to estimate $M$ from such experiments.
But we are optimistic that future developments in optogenetic tools will provide the necessary means to establish a more quantitative link between experimentally accessible quantities and modeling parameters.
}

\section*{Appendix}

Motile cells are \new{complex and} highly dynamic.
\old{and the design of mathematical models with moving boundaries is challenging.}
\new{
A fundamental challenge in the modeling of entire motile cells is the fact that, on a molecular level, many hundreds of interacting species are involved.
They are connected by a network of biochemical reactions and, in the case of cytoskeletal and membrane components, also contribute to the mechanics of the movement process.
While parts of this network have been studied in detail, such as, for example, the role of Ras and phosphoinositide signaling or the treadmilling dynamics of actin filaments, other components are only partly explored or not known at all~\citep{devreotes_excitable_2017}.
To date, a detailed mechanistic whole-cell motility model is therefore out of reach.}

\new{For this reason, many modeling efforts rely on phenomenological concepts that are not rooted in the details of the underlying molecular mechanisms.
They are built on a small number of fundamental physical properties to account for the observed macroscopic behavior.}
\new{Here, a widely used approach are reaction-diffusion models in combination with a dynamic phase field to describe the cell boundary~\citep{shao_computational_2010,aransonPhysicalModelsCell2016}.
Instead of a detailed mechanical model of the cell membrane and cortex, the phase field takes only the most basic physical properties into account that govern the dynamics of the cell border, namely, interfacial tension, volume conservation, and the presence of active forces that are locally exerted by the cytoskeletal activity.
Similarly, the vast network of interacting intracellular components and their subcellular transport is approximated by a small set of reacting and diffusing species (reaction-diffusion system) that captures the observed patterns of intracellular activity.
In most cases, these species do not correspond to a single molecular player.
They are effective lumped variables that represent an entire set of molecular components, such as, for example, typical cell front markers that trigger protrusive activity at the leading edge (active Ras, PIP$_3$, freshly polymerized actin).
They are coupled to the active tension in the phase field, so that cell shape changes may occur as a consequence of the intracellular dynamics.}
\new{Despite the obvious conceptional shortcomings of such simplistic models, they serve an important purpose.
They reveal the basic dynamical properties of the system and cast them into the form of mathematical equations, which can be analyzed numerically and, in some cases, even analytically to provide results that can be challenged by future experiments.}

\new{
Numerous research groups have successfully used this approach to describe the motility of different cell types, such as keratocytes, neutrophils, and {\it D.~discoideum} cells~\citep{shao_coupling_2012,najem_phase-field_2013,imotoComparativeMappingCrawlingcell2021}.
Also the collective behavior of ensembles of many moving cells~\citep{lober_collisions_2015,moure_phase-field_2019} and their interactions with soft substrates, solid boundaries, and microstructures have been studied using such phenomenological phase field models~\citep{lober_modeling_2014,kulawiak_modeling_2016,hondaMicrotopographicalGuidanceMacropinocytic2021}.
While three-dimensional versions of such models became recently available~\citep{cao_minimal_2019,winkler_confinement_2019}, most phase field models to date are designed to describe two-dimensional projections of a cell.}

\new{We have recently proposed a phase field model}
\old{Recently, we developed a reaction-diffusion model for the intracellular dynamics combined with a phase field} to account for cell shape changes and locomotion of \textit{D.~discoideum} cells to address questions of cell crawling strategies and cell-to-cell variability~\citep{alonso_modeling_2018,morenoModelingCellCrawling2020}.
Here, we propose an extension of this model to account for the experimental observations reported in this article.
\new{Similar to earlier models by others,}
\old{Specifically,} we consider \old{a} \new{the following} phase field equation for the cell shape,
\begin{eqnarray}
\tau \frac{\partial \phi}{\partial t} =\gamma \left(\nabla^2 \phi -\frac{G'(\phi)}{\epsilon^2}\right)
-\,\beta \left| \nabla \phi \right| 
\label{pf}
+\,\alpha\, \phi\, c \left| \nabla \phi \right| , 
\end{eqnarray}
where $\phi=1$ denotes the interior of the cell, $\phi=0$ the exterior, and $\tau$ the characteristic time scale of the contour dynamics.
The three terms on the right-hand side of Eq.~(\ref{pf}) represent surface tension, area conservation, and active tension, respectively, and will be briefly discussed in the following.

\new{The first term ensures that, due to the membrane tension $\gamma$, the cell shape will converge to a circle if no other forces are applied.}
\old{The magnitude of the first term is set by the surface tension parameter $\gamma$,}
\old{$\epsilon$ controls the width of the transition zone between the two phases (i.e. the width of the cell boundary), and}
\new{Here, the function} $G(\phi)$ is a double well potential \new{that defines the inside ($\phi=1$) and the outside ($\phi=0$) of the cell as the two stable states of the phase field variable.}
We use $G(\phi) = 18\phi^2 (1 - \phi)^2$, for details see~\citet{alonso_modeling_2018}.
\new{The parameter $\epsilon$ controls the width of the transition zone between the inside and outside, i.e., the width of the cell border.}

The second term enforces \new{a constant cell volume, i.e., a constant cell area in the case of a two-dimensional projection.}
\new{In our present simulations, the} area conservation \old{and} was modified compared to the earlier versions of our model.
Previously, the cell area was kept constant at all times.
However, in our experiments, we recorded a cross-sectional cell area that may strongly fluctuate due to three-dimensional deformations of the cell body.
For this reason we adapted the area conservation term to allow for fluctuations of the cell area around an average value $A_0$ with a characteristic time scale $\tau_{\beta}$,
\begin{equation}\label{eqbeta}
\tau_{\beta} \frac{\partial \beta }{\partial t} = -\beta + \beta_{c} \left(\int \phi\, dA - A_0 \right) \, ,
\end{equation}
where the parameter $\beta_c$ determines how tightly the cell area is regulated to match the \old{reference} \new{average} value $A_0$.
In the limit $\tau_\beta\rightarrow 0$, we recover the area conservation term of our previous model~\citep{alonso_modeling_2018,morenoModelingCellCrawling2020}.

\new{As a consequence of} \old{Depending on} intracellular processes, \new{in particular dynamical rearrangements of the actin cytoskeleton}, the cell may generate active tension that deforms the cell contour \new{to form, for example, pseudopodia.}
\new{These active tensions are} represented by the third term on the right-hand side of Eq.~(\ref{pf}).
The amplitude of the active tension is set by the parameter $\alpha$, and the intracellular components that actively drive membrane deformations are represented by the concentration $c$ of an effective force generating component.
\new{In a motile cell, the active tension depends on the cooperative action of a large number of interacting cytoskeletal components that are regulated by an upstream signaling network.
As many of these interactions and their governing rate constants are not known, the present phenomenological model lumps their joint action into a single effective activatory component $c$.
It represents the concentration of typical cell front markers, such as active Ras, PIP$_3$, or freshly polymerized actin, that are associated with regions of protrusive activity at the cell front.}
For the dynamics of $c$ we previously proposed a noisy bistable reaction-diffusion system that is confined inside the two-dimensional domain of the cell, and thus is coupled to the phase field in the following way~\new{\citep{alonso_modeling_2018,morenoModelingCellCrawling2020}},
\begin{equation}
\label{eq1}
\frac{\partial (\phi c) }{\partial t} = \phi [k_a\, c\,(1-c)(c-\delta(c)) - \rho\, c]
+\,F_R \,\phi\,\xi(x,t) + \nabla \left(\phi D \nabla  c\right) \, .  
\end{equation}
%
\new{A bistable system seems a natural choice to capture observations, where, due to the presence of waves, the bottom cortex of the cell shows coexisting regions of low and high F-actin concentration.}
\new{For the associated nonlinearity in the kinetics of $c$ several candidate processes have been identified, such as a positive feedback loop involving Ras, PI3K and F-actin~\citep{sasaki_g_2007}, or the autocatalytic enhancement of actin polymerization due to branching.} 
\new{At the whole cell level, the reaction rate $k_a$ controls the cell's speed, polarity, and persistence of motion, as has been systematically studied earlier~\citep{alonso_modeling_2018}.
The parameters $\rho$ and $D$ designate the degradation rate and the diffusion coefficient of $c$, respectively.}
\old{where} \new{For} the noise \old{follows} an Ornstein-Uhlenbeck dynamics \new{was chosen}, 
\begin{equation}\label{noise}
\frac{d \xi}{d t} = -k_{\eta}\, \xi + \eta\, ,
\end{equation}
\new{which means that temporal correlations in the fluctuations of $c$ decay exponentially with $k_{\eta}^{-1}$.}
Here, $\eta$ is a Gaussian white noise with zero mean
$\langle\eta\rangle=0$
and a variance of 
$\langle\eta(\mathbf{x},t)\eta(\mathbf{x'},t')\rangle=
2\sigma^2\delta(\mathbf{x}-\mathbf{x'})\delta(t-t')$.
In contrast to previous versions of the model~\citep{alonso_modeling_2018,morenoModelingCellCrawling2020}, the noise appears over the entire ventral surface of the cell~\citep{flemming_how_2020} and is modulated by a prefactor of $F_R=1.77(1-Q)^2$,
where $Q$ denotes the fraction of the cell area covered with the concentration $c$,
\begin{equation}\label{eqQ}
Q = \frac{\int \phi\, c\, dA }{\int \phi \, dA } \, .
\end{equation}
Here, $Q=1$ corresponds to a cell completely covered with a high level of $c$ and $Q=0$ indicates vanishing concentrations of $c$ across the cell area.
The prefactor $F_R$ ensures that the noise is reduced with decreasing concentration~$c$, and the value $F_R=1$ is recovered for $Q=0.25$ as previously employed \citep{morenoModelingCellCrawling2020}.

In the previous version of our model, the fraction of the cell area that is covered by \new{wave} patches of high values of $c$ was regulated to match a fixed value of $Q_0$ on average.
\new{This was achieved by dynamically tuning the quantity $\delta(c)$ in Eq.~(\ref{eq1}), which controls whether wave patches grow or shrink~\citep{alonso_modeling_2018}.}
Depending on the choice of $Q_0$, amoeboid (small $Q_0$) or fan-shaped cells (large $Q_0$) were observed in numerical simulations of the model~\citep{morenoModelingCellCrawling2020}.
However, the experimental observations reported in this article suggest that cells can spontaneously switch between both modes of locomotion.
We therefore extended the model such that $Q_0$ is no longer a fixed parameter but may switch in a noise induced fashion between small and large values, depending on the current relative coverage of the cell area with $c$ as given by the quantity $Q$,
\begin{equation}
\label{eq:cases}
    Q_0= 
\begin{cases}
    0.25, & \text{if } Q\leq Q_{Th}\\
    0.75, & \text{if } Q\geq Q_{Th} \, ,
\end{cases}
\end{equation}
where the parameter $Q_{Th}$ may be chosen between 0 and 1.
For large $Q_{Th}$ patches of high $c$ remain small, cells move in an amoeboid fashion, and switches to the fan-shaped state are rare.
In contrast, for small values of $Q_{Th}$ most of the cell area will be covered by high values of $c$ driving the cell into fan-shaped motion.
\new{To elucidate the role of the newly introduced threshold parameter $Q_{Th}$ for the switching process, we have systematically scanned $Q_{Th}$ between 0 and 1 in steps of 0.1, see Fig.~\ref{fig:6:phase_diagram}.}

In addition, the quantity $\delta(c)$ that determines whether patches of large $c$ are growing or shrinking is now dynamically adapting to changes in $Q_0$ with a characteristic time scale $\tau_\delta$ according to the following equation,
%
\begin{equation}\label{delta}
\tau_\delta \frac{\partial \delta (c) }{\partial t} =  -\delta(c) + \delta_0 + M \left( \int \phi\, c\, dA - Q_0  \int \phi\,dA  \right) \, .
\end{equation}
Here, $\delta_0=0.5$ corresponds to the neutral situation, where patches of $c$ are neither growing nor shrinking, and
the parameter $M$ determines how tightly the coverage with $c$ is regulated to the target value of $Q_0$.

\new{
For the simulations presented here, we used the previously established set of model parameters without any further adaptation to the present situation~\citep{morenoModelingCellCrawling2020}.
Only those parameters that are related to the extension of the model were newly chosen.
Specifically, the time scales of cell area and wave area variations, $\tau_{\beta}$ and $\tau_{\delta}$, were set to $0.2~s^{-1}$ to match the dynamics observed in our experiments.
The concentration constraint parameter $M$ and the threshold value $Q_{Th}$ were systematically changed to explore their impact on the switching behavior, see Fig.~\ref{fig:6:phase_diagram} and the corresponding part of the Discussion.
For the area constraint parameter $\beta_C$ we did not perform a systematic scan because the dynamics in the cell area fluctuations is not critical for the switching between amoeboid and fan-shaped cells.
All model parameters are displayed in Table~\ref{tab:1} in the Supplementary Material.
}


\section*{Supplemental Data}


\section*{Data Availability Statement}
The raw data supporting the conclusions of this article will be made available by the authors upon request.


\section*{Author Contributions}

TM, CB implemented research concept,
TM, CB, EM, SA interpreted results,
TM analyzed data,
SF contributed experimental data,
DS, MH, WH contributed data analysis tools,
SA designed modeling framework,
EM performed numerical simulations,
EM, DS, SF, SA contributed to drafting the manuscript,
CB, TM wrote original version the manuscript,
CB, SA designed research,
CB, SA, MH, WH supervised the project,
all authors contributed to manuscript revision, read, and approved the submitted version.



\section*{Acknowledgments}
We thank Kirsten Sachse for technical assistance. 


\section*{Funding}
The research of TM, DS, MH, WH, and CB has been partially funded by the Deutsche Forschungsgemeinschaft (DFG) – Project-ID 318763901 – SFB1294.
EM and SA acknowledge financial support by grant PGC2018-095456-b-I00 funded by MCIN/AEI/ 10.13039/501100011033 and by “ERDF A way of making Europe”, by the European Union.


\section*{Conflict of Interest Statement}

The authors declare that the research was conducted in the absence of any commercial or financial relationships that could be construed as a potential conflict of interest.


\bibliographystyle{Frontiers-Harvard} 
\bibliography{manuscript_bibliography_fixed.bib}



\end{document}


\onecolumn
\firstpage{1}

\title[Supplementary Material]{{\helveticaitalic{Supplementary Material}}}

\maketitle

\begin{figure*}[h!] 
\begin{center}
    \includegraphics[width=0.95\linewidth]{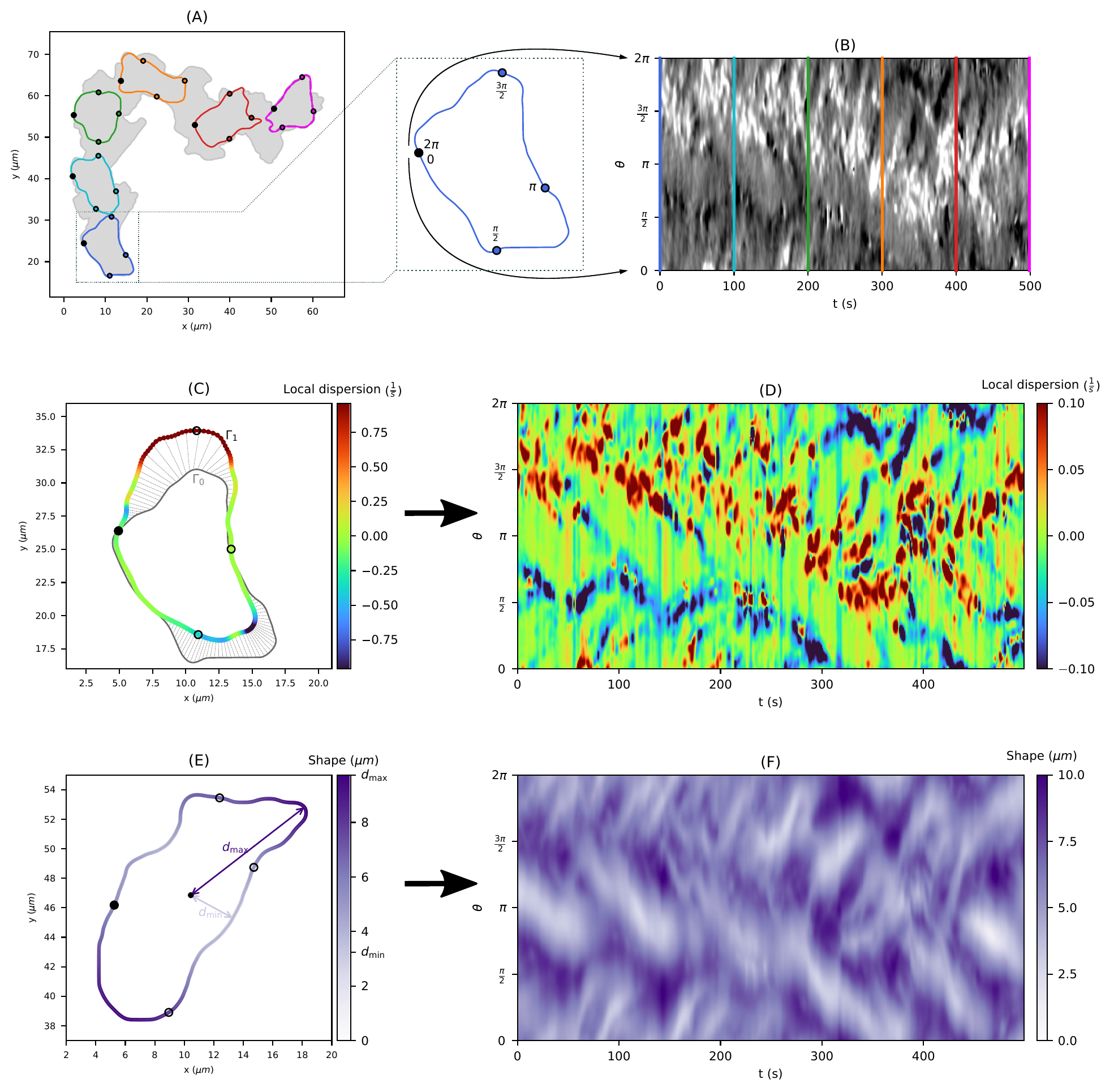}
\end{center}
\caption{
\new{
Computation of geometric and dynamic quantities of a motile cell and their representation as kymographs.
(A) Cell track with underlying spatio-temporal coordinate system shown for exemplary contours and virtual markers: $\theta=0$ (black filled circle) and $\theta\in\{\frac{\pi}{2},\pi,\frac{3\pi}{2}\}$ (empty circle).
(B) A quantity of interest (the example here is the local motion) can be displayed as a two-dimensional kymograph representation, with the time coordinate on the x-axis and the normalized arc length coordinate $\theta\in[0,2\pi)$ on the y-axis (colored lines representing single cell contours).
(C) Local dispersion showing regions of virtual marker thinning (expansions, red), virtual marker clustering (retractions, blue), and neutral regions (green).
(E) Distances from virtual markers on the contour to the center of mass of the cell (so-call contour shape, color-coded in purple).
(D, F) Corresponding kymographs showing local dispersion and distances to the center of mass as a function of time and space.
}
}
\label{fig:S1:explanation_kymographs}
\end{figure*}

\newpage

\begin{figure*}[h!] 
\begin{center}
    \includegraphics[width=0.80\linewidth]{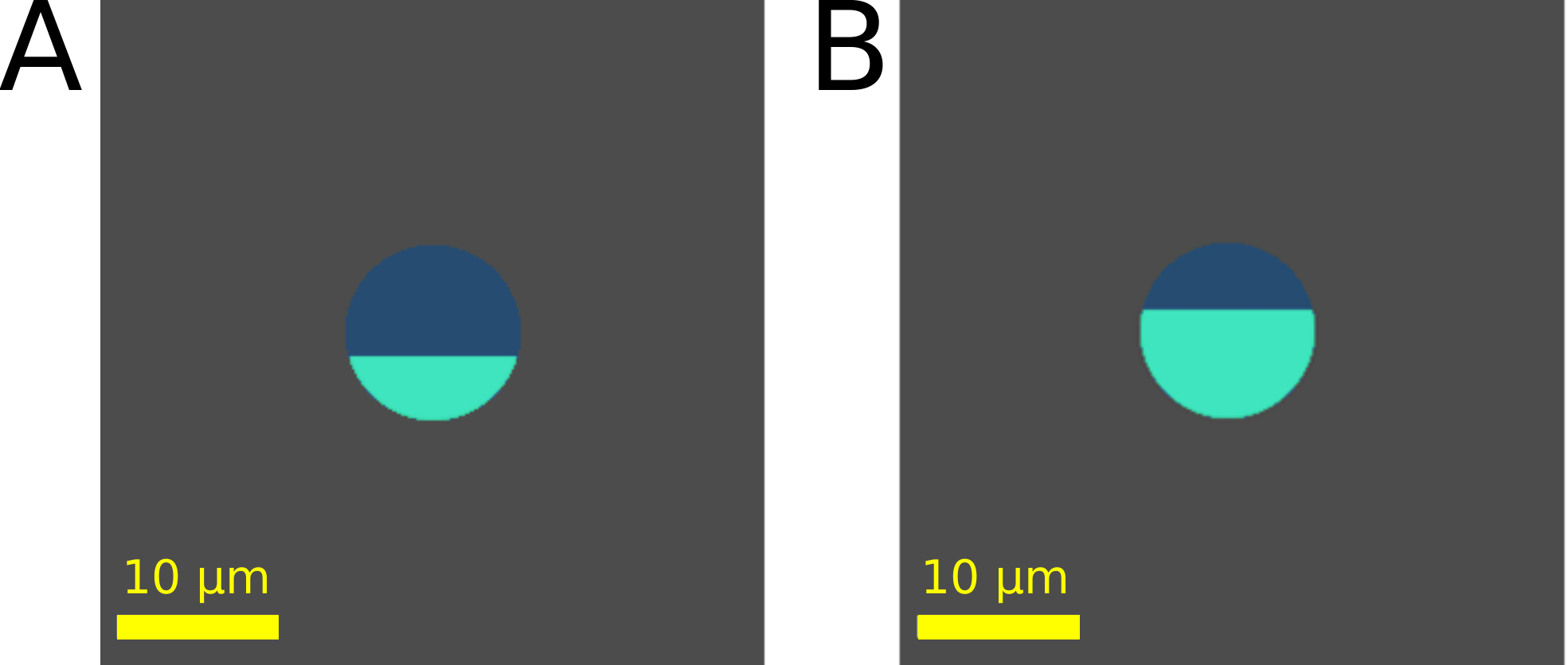}
\end{center}
\caption{
\new{
Initial conditions for the model simulations.
All simulations, which have been taken into account for the analysis in this article, started with either $25\%$ (A) or $75\%$ (B) of the cell area covered by high values of the effective force generating component $c$.
After starting the simulations, the cell shape and the intracellular distribution of $c$ randomly evolved according to the stochastic model equations, so that correlations with the initial cell shape and $c$ distribution were rapidly lost, see also the corresponding movie.
}
}
\label{fig:S2:initial_simulation_conditions}
\end{figure*}

\newpage

\begin{figure*}[h!] 
\centering    
    \includegraphics[width=.6\linewidth]{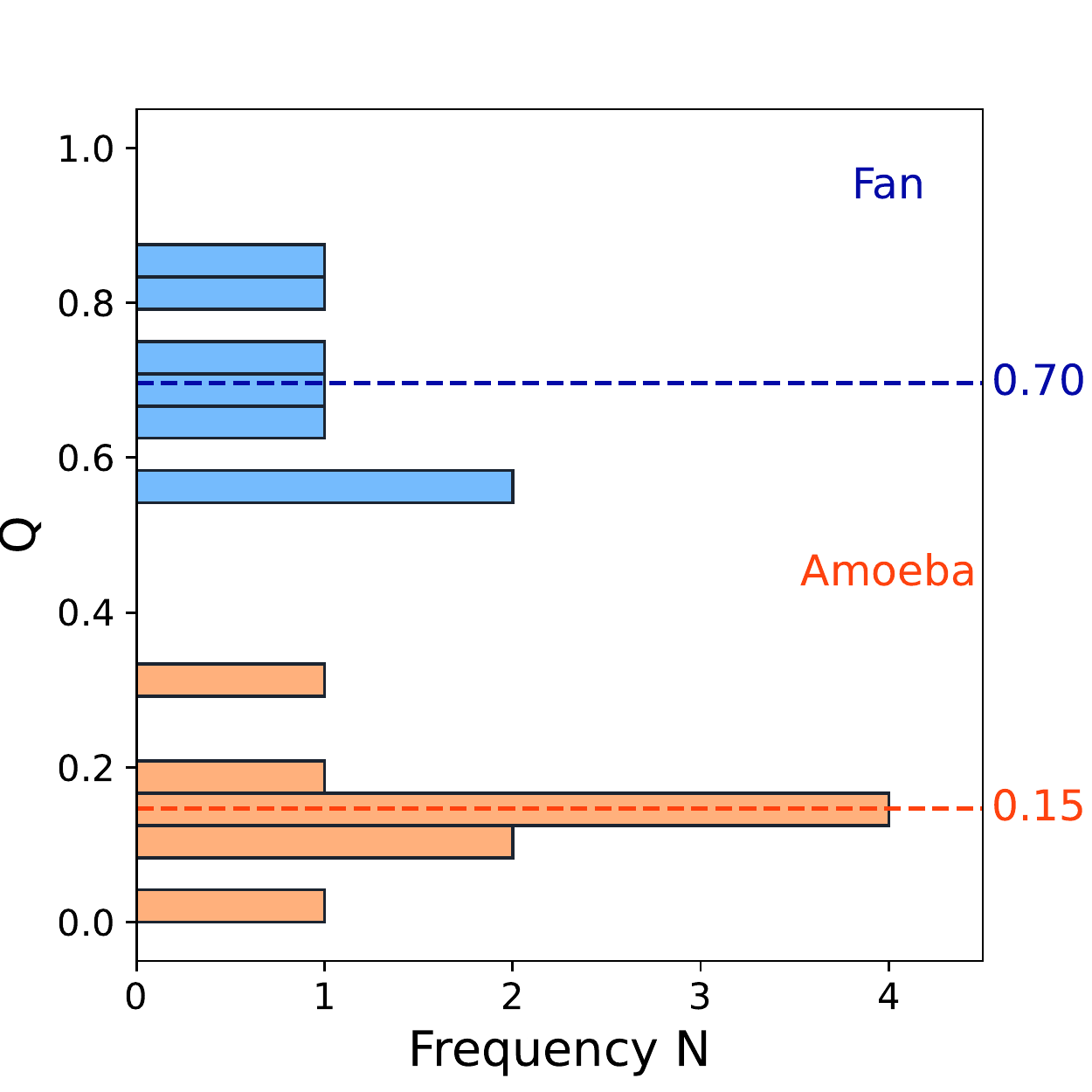}
\caption{
\new{
Histogram of $Q$ values generated for experimentally recorded cells taken from \cite{ghabache_coupling_2021} and \cite{miao_altering_2017}, see details below.
All cells were recorded by fluorescence microscopy.
The blue bars represent cells classified in the corresponding papers as fan-shaped cells.
The dark blue dashed line is the mean value $Q=0.70$ of all fan-shaped cells.
The orange bars represent cells classified in the corresponding papers as amoeboid cells.
The dark orange dashed line is the mean value $Q=0.15$ of all amoeboid cells.
The fact that all $Q$ values of amoeboid cells are below $Q=0.50$ and all $Q$ values of fan-shape cells are above $Q=0.50$ support our classification of amoeboid and fan-shape cells based on the value of~$Q$.
}
}
\label{fig:S3:comparison_of_Q_values}
\end{figure*}

\newpage

\new{
\subsection*{Small protrusions at the leading edge of fan-shaped cells}}
\label{sec:minipods}
%
\begin{figure*}
\centering
    \includegraphics[scale=0.55]{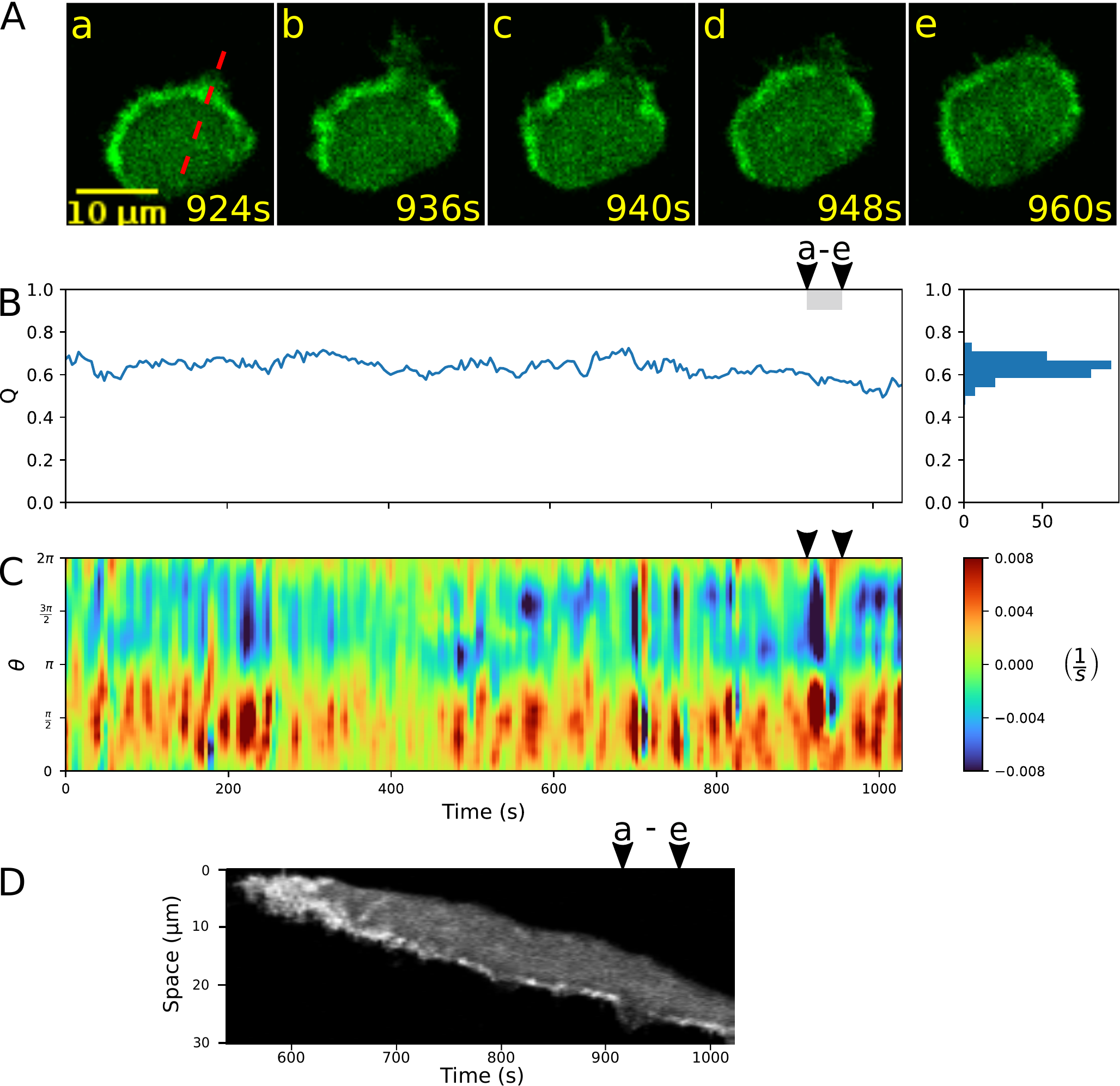}
    \hspace{3mm}
\caption{
\new{
    Small protrusions at the leading edge of fan-shaped cell.
    (A)~Snapshots displaying growth and decay of a protrusion at the leading edge~(a) to~(e).
    (B) Time evolution of the relative wave area~$Q$.
    The shaded area in between~(a) and~(e) indicates the episode of protrusion formation.
    Histogram on the right-hand side shows the frequency of $Q$-values.
    (C) Kymograph of the local dispersion taken along the cell contour.
    (D)~Fluorescence kymograph taken along the dashed red line displayed in the first panel of~(A).
    Arrowheads above the kymographs mark the time points of the corresponding fluorescence images in~(A).
    }
    \label{fig:S4:minipods_exp}}
\end{figure*}

\new{
At the leading edge of persistently moving fan-shaped cells, we repeatedly observed the emergence of small, short-lived protrusions.
In Fig.~\ref{fig:S4:minipods_exp}A, the formation and decay of such a protrusion is illustrated in a sequence of snapshots (a)--(e) from the fluorescence recording of a stable fan-shaped cell.
The entire event extends over a short time window of about 36~s, which is typical for the life-time of these structures.
Due to their small size, their formation does not affect the overall value of the relative wave area $Q$, see Fig.~\ref{fig:S4:minipods_exp}B, where $Q$ remains around $0.6$ for the entire measurement time.
However, in the local dispersion kymograph of the cell contour in panel C, protrusion formation can be clearly detected as a strongly localized, extending region (red), with a simultaneous increased retraction along the rest of the contour (blue), see the time period between (a) and~(e), as well as the time period shortly after $t=200$~s.
}

\new{
The growing protrusion is accompanied by a decrease of the fluorescence intensity in its vicinity.
In particular, the front of the actin wave that pushes the cell forward is directly affected by the growing protrusion, such that the leading wave segment is disrupted, indicating a depletion of F-actin, which causes a breakup of the wave segment at the position of protrusion formation, see Fig.~\ref{fig:S4:minipods_exp}A at time frames (b)--(d).
We thus conclude that the growing protrusion competes with the wave for the common pool of actin, consuming some of it in its surrounding in order to push the plasma membrane outwards. 
When the protrusion has decayed, the disrupted actin segment at the wave front heals and recovers its initial fluorescence intensity.
This can be also seen in the fluorescence intensity kymograph in Fig.~\ref{fig:S4:minipods_exp}D, taken along the red dashed line displayed in the first snapshot of panel~A.
}


\new{
Small protrusions at the leading edge of fan-shaped cells have been noted earlier~\citep{flemming_how_2020} and can be also induced by chemotactic cues~\citep{ecke_co-existence_2017}.
However, based on the limited amount of available data, their precise nature remains elusive.
They resemble small, short-lived pseudopods that are commonly observed in {\it D.~discoideum} cells when moving in an amoeboid fashion~\citep{swansonLocalSpatiallyCoordinated1982, rubinoLocationActinMyosin1984, wesselsCellMotilityChemotaxis1988} but it is also conceivable that their formation is driven by a blebbing mechanism~\citep{tyson_how_2014}.
}

\newpage

\begin{figure*}[h!] 
\begin{center}
    \includegraphics[width=0.80\linewidth]{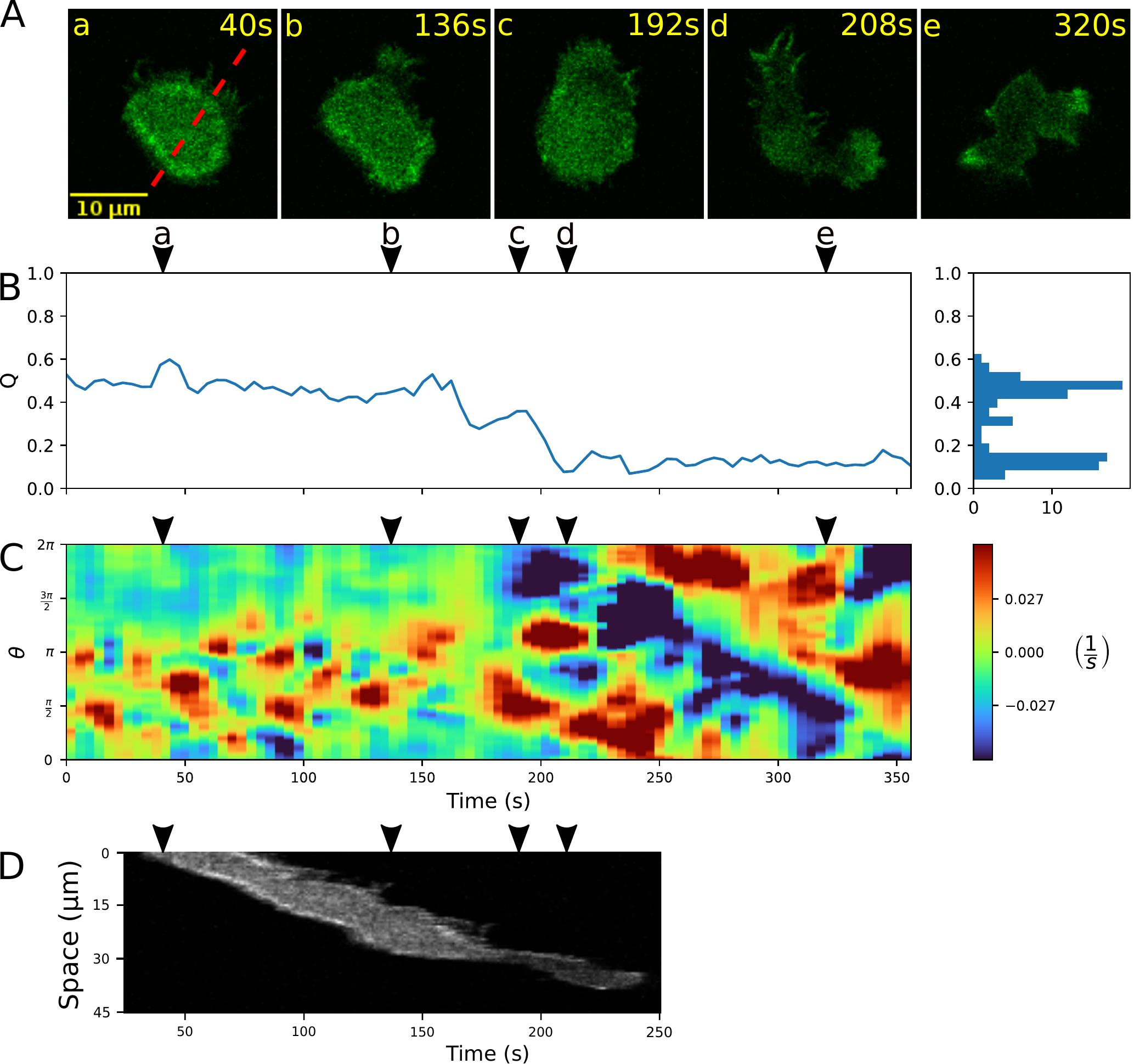}
\end{center}
\caption{
Breakdown of actin wave and simultaneous switching from fan-shaped to amoeboid mode of migration.
(A) Snapshot\new{s} \old{(a)} showing \old{the} \new{a} cell \new{in} moving fan-shaped \old{like} \new{mode} pushed by a stable actin wave \new{(a, b)}.
The breadown of the actin wave and the switching is visible \old{from} \new{in} (c) to (e)\old{. A} \new{, a}nd the cell moving in amoeboid mode of motion in (e).
(B) Time evolution of the relative wave area $Q$.
The arrowheads \old{corresponding to} \new{indicate the time points of} the snapshots in (A).
Histogram on the right-hand side shows the frequency of $Q$-values.
(C) Kymograph of the local dispersion taken along the cell contour.
(D) Fluorescence kymograph taken along the dashed red line indicated in the first panel of (A).
}
\label{fig:S5:breakdown_wave_exp}
\end{figure*}

\newpage

\begin{figure*}[h!] 
\begin{center}
    \includegraphics[width=.8\linewidth]{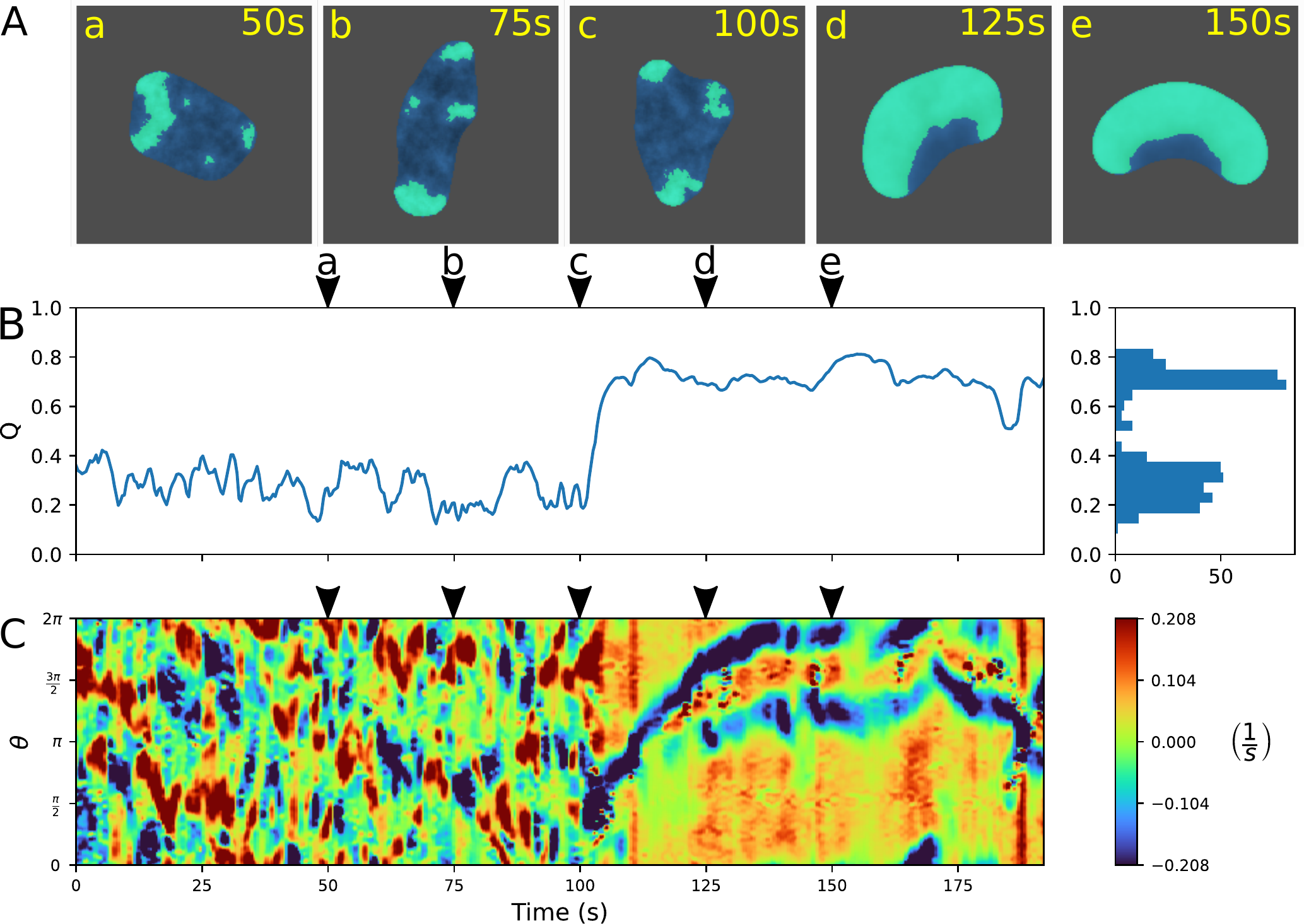}
\end{center}
\caption{
\old{One time s} \new{S}witching from amoeboid motion to fan-shaped migration mode in model simulations.
(A) Snapshots of amoeboid migration mode (a, b, c) switching to fan-shaped migration (d, e).
(B) Depiction of relative wave area \old{$Q=A_{C}/A_{W}$} \new{$Q=A_{W}/A_{C}$}, where at the left half\new{,} low values with high variations indicate the amoeboid migration\new{,} while at the right half high values of $Q$ with low variations are associated with fan-shaped motion.
(C) \old{The l} \new{L}ocal dispersion kymograph along the cell contour. 
The \old{dashed lines} \new{arrowheads} with label\new{s} \new{(}a\new{)} to \new{(}e\new{)} mark the time points corresponding to the snapshots in (A).
\new{The concentration constraint parameter was set to $M=0.00225$ and the threshold value to $Q_{Th}=0.5$;
for all other model parameters see Table~\ref{tab:1}.}
}
\label{fig:S6:switch_amoeba_fan_sim}
\end{figure*}

\newpage
\begin{figure*}[h!] 
\centering    
    \includegraphics[width=.8\linewidth]{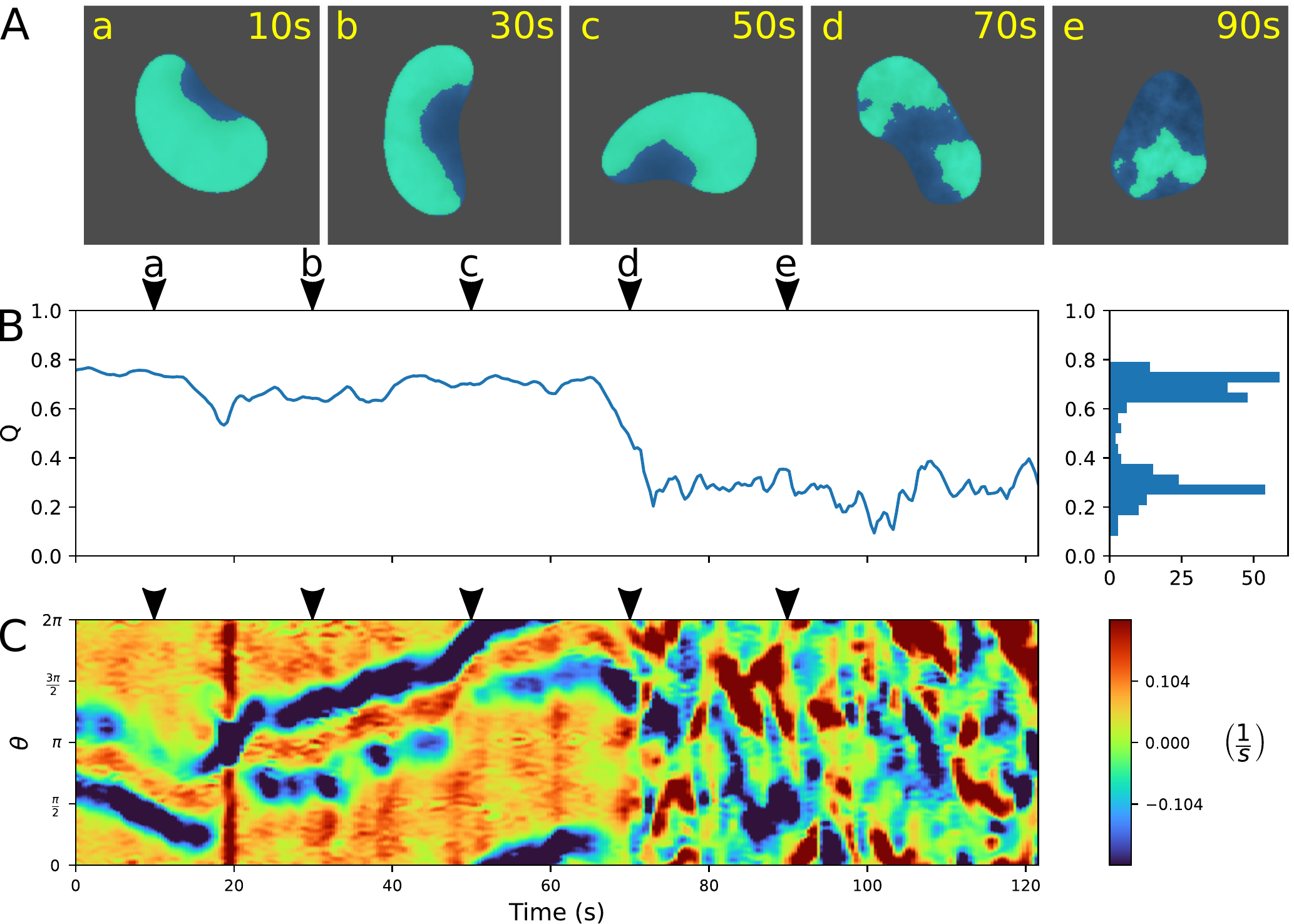}
\caption{
Model simulation of wave breakdown and associated motility mode switch\new{ing} from fan-shaped to amoeboid motion.
(A) Snapshots of fan-shaped mode (a, b, c) switching to amoeboid motility (d, e).
(B) Time trace of relative wave area \old{$Q=A_{C}/A_{W}$} \new{$Q=A_{W}/A_{C}$}.
\old{At} \new{In} the first half\new{,} high values with low variations are associated with the fan-shaped migration  mode\new{,} while \old{at} \new{in} the second half\new{,} low values with stronger variations indicate the amoeboid motion.
(C) \old{The l} \new{L}ocal dispersion kymograph along the cell contour. 
The \old{dashed lines} \new{arrowheads} with label\new{s} \new{(}a\new{)} to \new{(}e\new{)} mark the time points corresponding to the snapshots in (A).
\new{The concentration constraint parameter was set to $M=0.0015$ and the threshold value to $Q_{Th}=0.5$;
for all other model parameters see Table~\ref{tab:1}.}
}
\label{fig:S7:breakdown_wave_sim}
\end{figure*}

\newpage
\begin{figure*}[h!] 
\centering    
    \includegraphics[width=.8\linewidth]{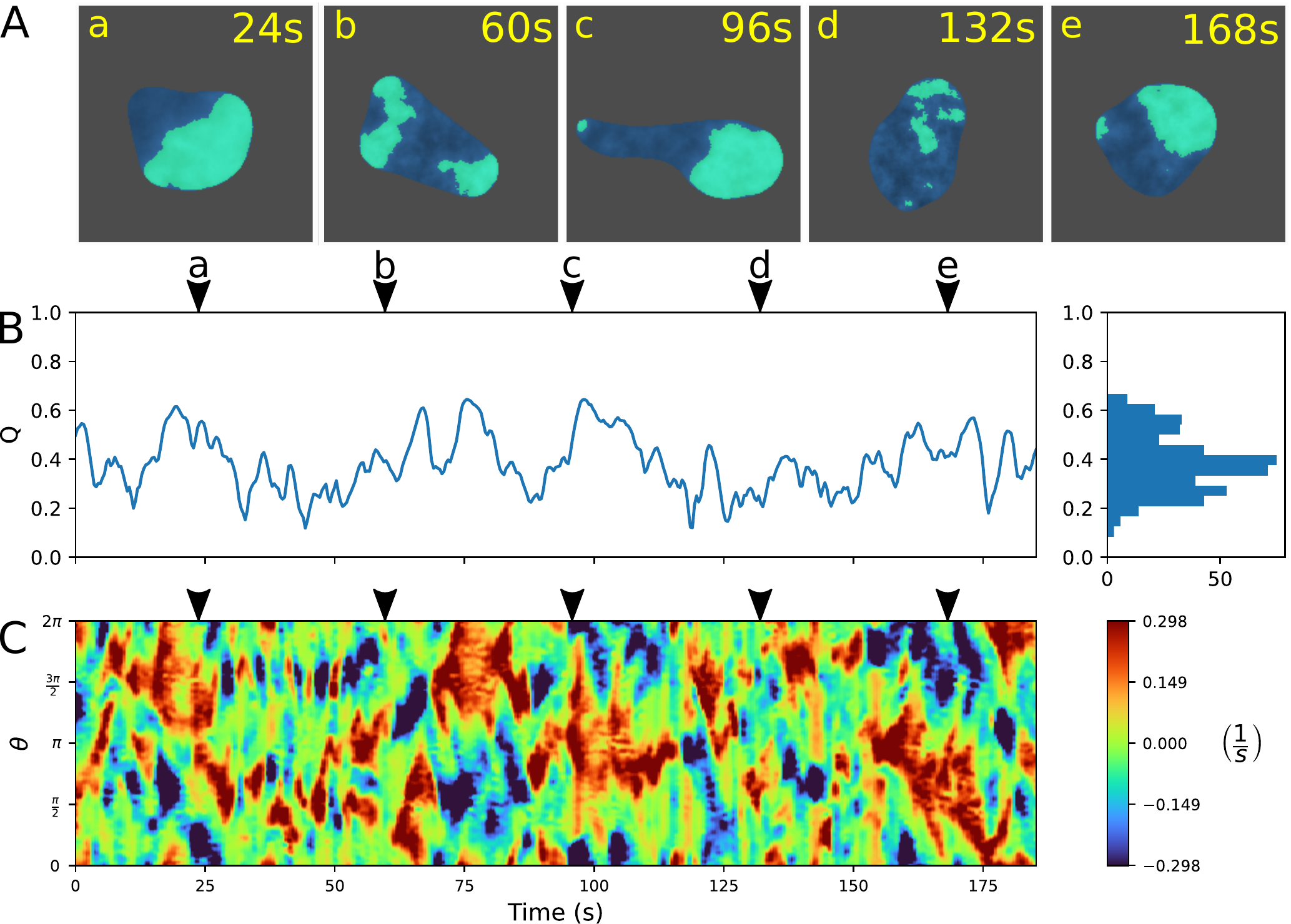}
\caption{
Model simulation of failed attempt\old{s} to switch migration mode from amoeboid to fan-shaped.
(A) Snapshots of growing and decaying waves which did not \old{become stable} \new{switch to a stable fan-shaped configuration}.
(B) Depiction of relative wave area \old{$Q=A_{C}/A_{W}$} \new{$Q=A_{W}/A_{C}$}.
The \old{high} \new{strongly} fluctuating $Q$ value with a mean around $0.4$ is \old{representing} \new{reflecting} the \old{consistently} \new{repeatedly} growing waves which did not archive stability.
(C) \old{The l} \new{L}ocal dispersion kymograph along the cell contour. 
The \old{dashed lines} \new{arrowheads} with label\new{s} \new{(}a\new{)} to \new{(}e\new{)} mark the time points corresponding to the snapshots in (A).
\old{In contrast to the local dispersion kymographs about no distinct pattern corresponding to one of the to migration modes could be found.}
\new{The concentration constraint parameter was set to $M=0.0045$ and the threshold value to $Q_{Th}=0.4$;
for all other model parameters see Table~\ref{tab:1}.}
}
\label{fig:S8:failed_switch_sim}
\end{figure*}

\newpage

\new{
\subsection*{Analysis of the distance from the contour to the center of the cell}}
\label{sec:centroid_to_contour_comparison}
\new{
Our model was not specifically tuned to reproduce geometrical features of the experimentally observed cell shapes.
Thus, the distance from the center of the cell to the contour provides an additional opportunity to compare experimental observations and simulation results, i.e., an independent quality check for the performance of our model.
}

\begin{figure*}[h!] 
\centering    
    \includegraphics[width=1.00\linewidth]{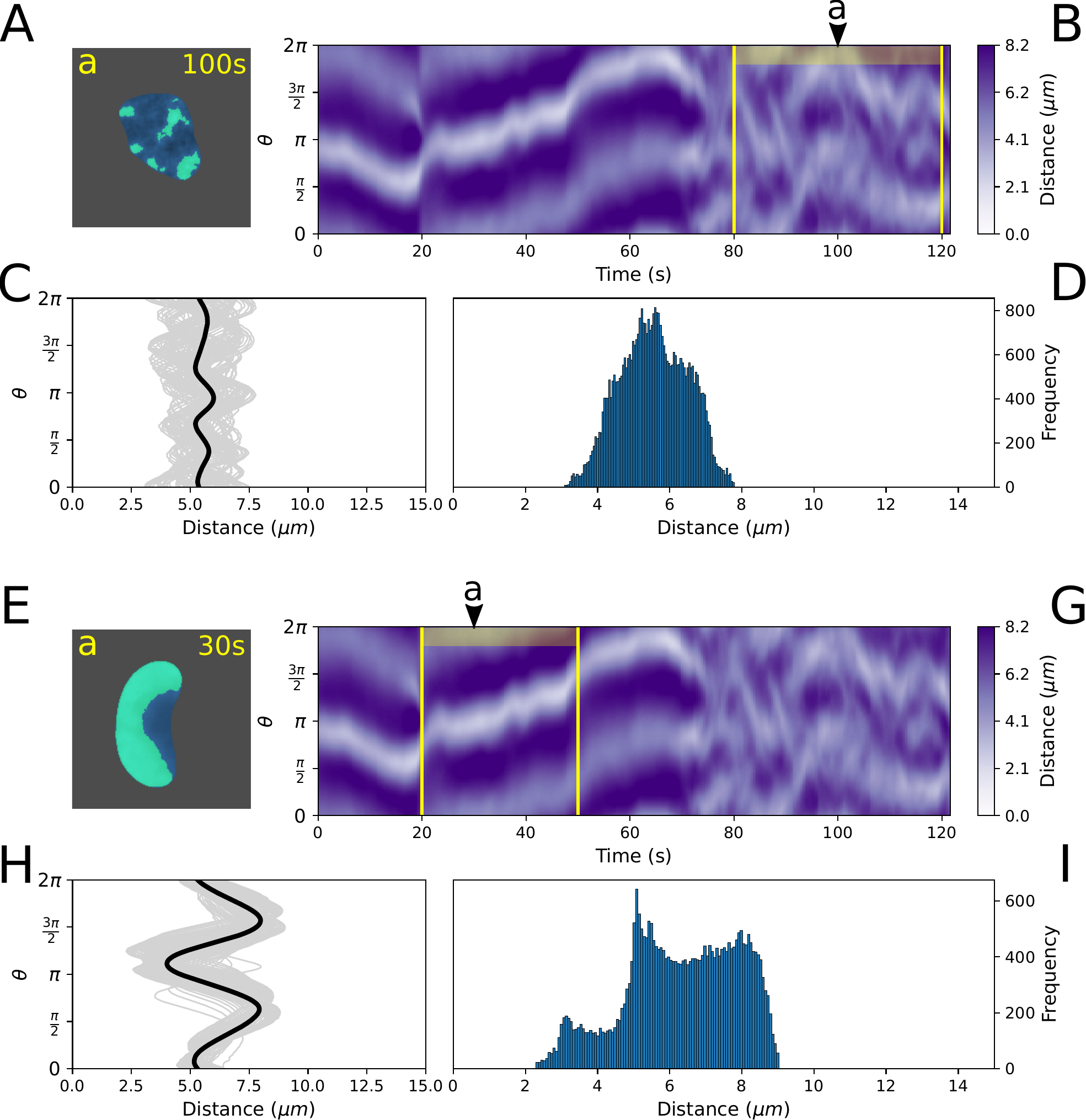}
\caption{
\new{
Illustration of the cell center to contour distance for a numerically simulated cell while moving in the amoeboid (A~--~D) and in the fan-shaped mode (E~--~I).
(A, E) Snapshots of the cell while migrating in amoeboid (A) and fan-shape mode (E).
(B, G) Kymograph of the cell center to contour distance, illustrating the distance of each position on the contour ($0$ to $\pi$) over time.
White indicates small and dark blue large distances of a contour point from the center.
The arrowheads (a) mark the corresponding time points of the cell shown in the snapshots (A) and (E), respectively.
For further explanations of the kymograph concept see Fig.~\ref{fig:S1:explanation_kymographs}.
(C, H) Distance to center plot of each individual contour (gray lines) in the time interval highlighted in yellow in (B, G).
The black line is the average of all contours in the highlighted time interval.
(D, I) Histogram of the distance to the center for all positions on the contour over the highlighted time interval.
}
}
\label{fig:S10:distance_to_center_sim}

\end{figure*}

\new{
In order to interpret the data, let us first consider what we expect for the different cases.
During amoeboid migration without any external cue, pseudopod activity is randomly distributed all around the cell contour.
Hence, on a time scale much longer than the typical pseudopod lifetime, the distance from the center to the contour of an amoeboid cell will evolve randomly and will, when averaged over time, approach half of the cell diameter for each point on the contour (a typical cell diameter is about $10\mu$m).
The mean distance to center for amoeboid contours will thus converge to a flat profile at a distance of around $5\mu$m, while the histogram of the center to contour distances for all individual contour positions will be symmetrically distributed around this value.
}
\new{
This is indeed observed both in the experimental (Fig.~\ref{fig:S9:distance_to_center_exp}) and in the numerical data (Fig.~\ref{fig:S10:distance_to_center_sim}).
When considering time intervals, during which the cell moves in an amoeboid fashion (highlighted in yellow in the kymographs (B) in Figs.~\ref{fig:S9:distance_to_center_exp} and \ref{fig:S10:distance_to_center_sim}), we see in the experimental as well as in the simulation data no distinct coherent patterns.
Figures~\ref{fig:S9:distance_to_center_exp}~(C) and \ref{fig:S10:distance_to_center_sim}~(C) depict the distance from the center to the contour for all individual contours (grey lines) in the highlighted time interval, and the black line is the average over all these contours, displaying an almost constant value of around $5\mu$m.
Figures~\ref{fig:S9:distance_to_center_exp}~(D) and \ref{fig:S10:distance_to_center_sim}~(D) show the histograms for all contour positions during the highlighted time interval, which almost symmetrically center around the mean value.
}

\new{
In the case of fan-shaped motion, the cell adopts a kidney-like shape, elongated perpendicular to the direction of motion.
Thus, the front and rear parts of the cell border are closer to the center compared the left and right parts.
This shape is stably maintained while the cell is moving.
Hence, we expect to see a characteristic pattern in the distances of the contours from the center for these cells.
This is indeed observed during time intervals, when cells move in the fan-shaped mode, 
see the time interval highlighted in yellow in the kymographs (G) in Figs.~\ref{fig:S9:distance_to_center_exp} and \ref{fig:S10:distance_to_center_sim}.
Here, the kidney shape is reflected by an alternating pattern of dark and bright stripes. 
This is also visible in the plots of the individual contours from the highlighted time interval (grey lines) and in their average (black line) in Figs.~\ref{fig:S9:distance_to_center_exp}~(H) and \ref{fig:S10:distance_to_center_sim}~(H).
As the kidney shape persists over time, the histograms of the distances taken over the individual contours display a characteristic asymmetric distribution with three local maxima, which is, however, more clearly seen in the numerical than in the experimental data, see Figs.~\ref{fig:S9:distance_to_center_exp}~(I) and \ref{fig:S10:distance_to_center_sim}~(I).
}

\begin{figure*}[h!] 
\centering    
    \includegraphics[width=1.00\linewidth]{figuresfig_S9_distance_to_center_exp.pdf}
\caption{
\new{
Illustration of the cell center to contour distance for an experimentally observed cell while moving in the amoeboid (A~--~D) and in the fan-shaped mode (E~--~I).
(A, E) Snapshots of the cell, taken by fluorescence microscopy, while migrating in amoeboid (A) and fan-shape mode (E).
(B, G) Kymograph of the cell center to contour distance, illustrating the distance of each position on the contour ($0$ to $\pi$) over time.
White indicates small and dark blue large distances of a contour point from the center.
The arrowheads (a) mark the corresponding time points of the cell shown in the snapshots (A) and (E), respectively.
For further explanations of the kymograph concept see Fig.~\ref{fig:S1:explanation_kymographs}.
(C, H) Distance to center plot of each individual contour (gray lines) in the time interval highlighted in yellow in (B, G).
The black line is the average of all contours in the highlighted time interval.
(D, I) Histogram of the distance to the center for all positions on the contour over the highlighted time interval.
}
}
\label{fig:S9:distance_to_center_exp}
\end{figure*}


\newpage

\clearpage

\mytextbox[0]{Video Figure \ref{fig:2:several_switches_exp}: Video of \new{a} fluorescent \dd{} DdB NF1 KO cell\old{s}. \old{Without any treatment the cell is} \new{Repeated} switching \old{the} \new{between amoeboid and fan-shaped} mode\new{s} of migration\new{.} \old{several times during locomotion. An amoeboid migration mode governed by pseudopods protruding outwards and retracting. And the fan-shaped migratory mode with it persistent motion driven by the actin wave pushing the cell.} The frame rate is 4 seconds. F-actin is labeled with Lifeact-GFP. Total time of video 1800 seconds.}

\mytextbox[0]{Video Figure \ref{fig:3:failed_switch_exp}: Video of \new{a} fluorescent \dd{} DdB NF1 KO cell\old{s}. Growth and decay of an actin wave is shown\new{, that does not stabilize into a fan-shape mode.} \old{Starting from a local focii the actin wave growth until the entire cell is filled. By doing so the actin wave overwrites the amoeboid cell shape and pushes the cell membrane into a fan-shaped form. Subsequently the actin waves pass through a rapid breakdown and decays.}  The frame rate is 4 seconds. F-actin is labeled with Lifeact-GFP. Total time 1188 seconds\new{.}}

\mytextbox[0]{Video Figure \ref{fig:4:switch_amoeba_fan_exp}: Video of \new{a} fluorescent \dd{} DdB NF1 KO cell\old{s}. A cell switching \old{one time} from amoeboid motion into the fan-shaped mode. The frame rate is 4 seconds. F-actin is labeled with Lifeact-GFP. Total time 1752 seconds\new{.}}

\mytextbox[0]{\new{Video Figure \ref{fig:S2:initial_simulation_conditions}: Model simulation video, showing the initial conditions for the model simulations.}}

\mytextbox[0]{Video Figure \ref{fig:S4:minipods_exp}: Video of \new{a} fluorescent \dd{} DdB NF1 KO cell\old{s}. The growth and decay of a small protrusion at the leading edge of a fan-shaped cell. The frame rate is 4 seconds. F-actin is labeled with Lifeact-GFP. Total time 104 seconds\new{.}}

\mytextbox[0]{Video Figure \ref{fig:S5:breakdown_wave_exp}: Video of \new{a} fluorescent \dd{} DdB NF1 KO cell\old{s}. Breakdown of the actin wave and the associated switching from fan-shaped migration to amoeboid motion. The frame rate is 4 seconds. F-actin is labeled with Lifeact-GFP. Total time 360 seconds\new{.}}

\mytextbox[0]{Video Figure \ref{fig:5:several_switches_sim}: Model simulation video\old{. S} \new{, s}howing \old{the bi-stable} \new{repeated noise-induced} switching \old{depending on the model noise.} \new{between amoeboid and fan-shaped mode.}} 

\mytextbox[0]{Video Figure \ref{fig:S6:switch_amoeba_fan_sim}: Model simulation video\old{. S} \new{, showing s}witch\new{ing} from amoeboid to fan-shaped motion.} 

\mytextbox[0]{Video Figure \ref{fig:S7:breakdown_wave_sim}: Model simulation video\old{. B} \new{, showing b}reakdown of the wave and \old{hence} switch\new{ing} from fan-shaped mode to amoeboid locomotion.}

\mytextbox[0]{Video Figure \ref{fig:S8:failed_switch_sim}: Model simulation video\old{. S} \new{, showing s}everal \old{failing}\new{failed} attempts to switch migration mode.}

\mytextbox[0]{\new{Figure \ref{fig:S3:comparison_of_Q_values}: 
Cells used for the analysis displayed in Fig.~\ref{fig:S3:comparison_of_Q_values} were taken from \citet{ghabache_coupling_2021} and \citet{miao_altering_2017}.
All cells were shown in the main figures of the articles, except two of them that were part of the supplementary material.
The analysis was performed on the fluorescence movies corresponding to the cells shown in the figures.
All movies were provided in the supplementary material of the articles.
}

\noindent\new{Cells classified as fan-shape cells in \citet{ghabache_coupling_2021}: Fig:2A (left side) - Movie EV1, Fig:2A (right side) - Movie EV2, Fig:2C (left side) - Movie EV1, and Fig:2C (right side) - Movie EV2.
}

\noindent\new{Cells classified as fan-shape cells in \citet{miao_altering_2017}: Fig:2D (right side) - Movie 41556 2017 BFncb3495 MOESM69 ESM, Fig:4A (middle) - Movie 41556 2017 BFncb3495 MOESM69 ESM, Fig:4A (middle) -Movie 41556 2017 BFncb3495 MOESM73 ESM, and Fig:4D (rigth side) 41556 2017 BFncb3495 MOESM74 ESM.
}

\noindent\new{Cells classified as amoeboid cells in \citet{ghabache_coupling_2021}: Fig:3A - Movie EV5, Fig:4A - Movie EV7, Movie EV6, and Movie EV8.
}

\noindent\new{Cells classified as amoeboid cells in \citet{miao_altering_2017}: Fig:2A (right side) - Movie 41556 2017 BFncb3495 MOESM68 ESM, Fig:2A (left side) - Movie 41556 2017 BFncb3495 MOESM68 ESM, Fig:2B (top left side) - Movie 41556 2017 BFncb3495 MOESM69 ESM, Fig:4A (left side) - Movie 41556 2017 BFncb3495 MOESM73 ESM, and Fig:4B (top left side) - Movie 41556 2017 BFncb3495 MOESM74 ESM.
}
}

\clearpage
\renewcommand{\arraystretch}{1.5}
\begin{table}[H]
\caption{Parameter values for the numerical model.}\label{tab:1}
	\begin{tabular}{@{}*{5}{l}}
		Parameter  & Value   & Units   & Meaning  & Reference\\
		\hline
		$D$        &   0.5  & $\mu m^2 /s$  &  Diffusion coefficient & \citep{alonso_modeling_2018}\\
		$k_a$      &   2    & $s^{-1}$  &  Reaction rate   & \citep{alonso_modeling_2018} \\
		$\rho$     &   0.02   & $s^{-1}$  &  Degradation rate	 	& \citep{morenoModelingCellCrawling2020}\\
		$\sigma$   &   0.15    &  $s^{-2}$ &  Noise strength	      & \citep{alonso_modeling_2018}\\
		$\tau$     &   2   & $pN s\,\mu m^{-2} $ & Membrane dynamics time-scale  & \citep{shao_computational_2010}\\ 
		& & & & \quad\&\,\citep{ziebert_model_2012}\\ 
		$\gamma$   &   2   & $pN$ & Surface tension	& \citep{shao_computational_2010}     	\\
		$\epsilon$ &   0.75   & $\mu m$ & Membrane thickness   & \citep{shao_computational_2010}\\
		$\beta$    &   22.22   & $pN \mu m^{-3} $ & Parameter for total area constraint & \citep{alonso_modeling_2018} \\
		$A_0$ 	   &   113   & $\mu m^{2}$ &   Area of the cell & \citep{alonso_modeling_2018}  \\
		$\delta_0$ & 0.5   & - &   Bistability critical parameter & \citep{alonso_modeling_2018}\\

		$k_{\eta}$ & 0.1   &  $s^{-1}$   &   Ornstein-Uhlenbeck rate & \citep{alonso_modeling_2018}\\
		$\alpha$   & 3 & $pN \mu m^{-1}$ & Active tension  & \citep{ziebert_model_2012}\\ 
		$\beta_C$    &  0.022    & $pN \mu m^{-5} $ & Control parameter for total area constraint &  \\
		$\tau_\beta$    &   0.2   & $ s $ & Time-scale for area constraint &  \\
		$\tau_\delta$    &   0.2   & $s$  & Time scale for bistability condition &  \\
	\hline	
	\end{tabular}
\end{table}

\bibliographystyle{Frontiers-Harvard} 

\bibliography{manuscript_bibliography_fixed.bib}